\def\readableformat{}
\ifdefined\readableformat
  \documentclass[twocolumn,longbib]{aastex701}
\else\ifdefined\showauthors
  \documentclass[manuscript,linenumbers,longbib]{aastex701}
\else
  \documentclass[manuscript,linenumbers,longbib,anonymous]{aastex701}
\fi
\fi

\usepackage{amsmath}
\usepackage{booktabs}
\usepackage{tabularx}
\usepackage{xcolor}
\usepackage{pict2e}

\newcommand{\kms}{\mathrm{km\,s^{-1}}}

\newcommand{\microm}{\mu\mathrm{m}}
\newcommand{\figpanel}[1]{\textcolor{white}{\textbf{#1}}}
\newcommand{\meteorpositionrow}[3]{%
  \begingroup
  \setlength{\unitlength}{\linewidth}%
  \begin{picture}(1,0.449)
    \put(0,0){\includegraphics[width=\linewidth]{#1}}%
    \put(0.065,0.369){\figpanel{#2}}%
    \put(0.565,0.369){\figpanel{#3}}%
  \end{picture}%
  \endgroup
}
\newcommand{\figureprovenance}[1]{\par\footnotesize\textit{Figure script: \texttt{\detokenize{#1}}.}}
\newcommand{\hidefigureprovenance}{\renewcommand{\figureprovenance}[1]{}}
\hidefigureprovenance
\graphicspath{{./}}
\providecommand{\micron}{\mu\mathrm{m}}

\begin{document}
\submitjournal{PSJ}
\shorttitle{PANSY Meteor Orbit Catalogue}
\shortauthors{Vierinen et al.}
\title{The PANSY Meteor Head-echo Orbit Catalogue: Continuous Antarctic Radar Observations of Southern Meteoroid Streams}

\author[orcid=0000-0001-7651-708X]{Juha Vierinen}
\affiliation{Department of Physics and Technology, UiT The Arctic University of Norway, Postboks 6050 Langnes, 9037 Troms\o, Norway}
\affiliation{Leibniz Institute of Atmospheric Physics at the University of Rostock, Schlo{\ss}stra{\ss}e 6, 18225 K\"uhlungsborn, Germany}
\email[show]{juha-pekka.vierinen@uit.no}
\correspondingauthor{Juha Vierinen}

\author[orcid=0000-0002-2364-8892]{Jorge L. Chau}
\affiliation{Leibniz Institute of Atmospheric Physics at the University of Rostock, Schlo{\ss}stra{\ss}e 6, 18225 K\"uhlungsborn, Germany}
\email{chau@iap-kborn.de}

\author[orcid=0000-0001-8824-9844]{Koji Nishimura}
\affiliation{Research Institute for Sustainable Humanosphere, Kyoto University, Gokasho, Uji, Kyoto 611-0011, Japan}
\email{nishimura@rish.kyoto-u.ac.jp}

\author[orcid=0000-0002-7082-9530]{Taishi Hashimoto}
\affiliation{National Institute of Polar Research, 10-3 Midori-cho, Tachikawa, Tokyo 190-8518, Japan}
\affiliation{Polar Science, Graduate Institute for Advanced Studies, SOKENDAI, 10-3 Midori-cho, Tachikawa, Tokyo 190-8518, Japan}
\email{hashimoto.taishi@nipr.ac.jp}

\author{Nico Pfeffer}
\affiliation{Leibniz Institute of Atmospheric Physics at the University of Rostock, Schlo{\ss}stra{\ss}e 6, 18225 K\"uhlungsborn, Germany}
\email{pfeffer@iap-kborn.de}

\author[orcid=0000-0002-6371-1016]{Daniel Kastinen}
\affiliation{Swedish Institute of Space Physics, Box 812, SE-981 28 Kiruna, Sweden}
\email{daniel.kastinen@irf.se}

\author[orcid=0000-0002-4257-4235]{Devin R. Huyghebaert}
\affiliation{Leibniz Institute of Atmospheric Physics at the University of Rostock, Schlo{\ss}stra{\ss}e 6, 18225 K\"uhlungsborn, Germany}
\email{huyghebaert@iap-kborn.de}

\author[orcid=0000-0002-2177-6751]{Johan Kero}
\affiliation{Swedish Institute of Space Physics, Box 812, SE-981 28 Kiruna, Sweden}
\email{johan.kero@irf.se}

\author[orcid=0000-0003-0113-8311]{Masaki Tsutsumi}
\affiliation{National Institute of Polar Research, 10-3 Midori-cho, Tachikawa, Tokyo 190-8518, Japan}
\affiliation{Polar Science, Graduate Institute for Advanced Studies, SOKENDAI, 10-3 Midori-cho, Tachikawa, Tokyo 190-8518, Japan}
\email{tutumi@nipr.ac.jp}

\author[orcid=0000-0002-6225-6066]{Kaoru Sato}
\affiliation{Department of Earth and Planetary Science, Graduate School of Science, The University of Tokyo, 7-3-1 Hongo, Bunkyo-ku, Tokyo 113-0033, Japan}
\email{kaoru@eps.s.u-tokyo.ac.jp}

\begin{abstract}
Meteor head echoes from high-power, large-aperture radars provide pulse-resolved positions and velocities for individual micrometeoroids in a submillimeter-radius range that contribute significantly to the mass influx to Earth. We present the first meteor head-echo orbit catalogue from the Antarctic Syowa Mesosphere--Stratosphere--Troposphere/Incoherent Scatter radar (PANSY). The catalogue contains two million meteors and 50 million pulse-resolved measurements. The observed radiant distribution contains the helion, antihelion, south toroidal and apex sources, and provides unprecedented head-echo coverage of southern ecliptic latitudes. The initial detection-height distribution is found to be double-banded, with both bands increasing in height with meteor speed and exhibiting distinct radiant distributions, consistent with a mixture of meteoroid-size differences and differential ablation. Estimation of dynamic mass indicates that the survey is sensitive to initial radii of approximately 100~$\mu$m. As part of an initial exploration of the catalogue, we investigate the nighttime $\alpha$ Capricornids (CAP) and the extended Daytime Capricornids-Sagittariids (DCS) radiant. Their similar activity durations at opposite nodes are consistent with membership in the CAP--169P/NEAT complex. The radiant distribution suggests a new meteor shower candidate, with peak flux near $\lambda_\odot=110^\circ$, mean Sun-centered ecliptic radiant $\lambda'_g=291.0^\circ,\beta_g=-48.2^\circ$, and $v_g=48.7\,\kms$. During catalogue production, event-level raw voltage cuts are retained temporarily to support improvements in data analysis. This first catalogue release, covering 2025 January 26 to 2026 July 26, will be useful for further studies of the southern-hemisphere Earth-crossing meteoroid population.
\end{abstract}


\section{Introduction}
\label{sec:introduction}

Meteoroids connect the interplanetary dust population to its cometary and asteroidal sources and deliver material to Earth's atmosphere. Their atmospheric trajectories constrain the radiant and speed distributions, meteoroid streams, and heliocentric orbits of the incident population \citep{ceplecha1998meteor,wiegert2009dynamical,soja2019imem2,nesvorny2011dynamical}. High-power, large-aperture radars detect the compact plasma surrounding an ablating meteoroid as a meteor head echo \citep{dimant2017formation}. Unlike specular trail echoes, a head echo can provide range, Doppler shift, and interferometric direction on every transmitted pulse, allowing a three-dimensional trajectory and orbit to be estimated for an individual meteoroid \citep{sato2000mu,chau+woodman-2004,schult2013determination,kastinenRadarAnalysisAlgorithm2022}. HPLA observations have previously resolved established meteor showers and compact stream radiants at Jicamarca, the MU radar, and MAARSY \citep{chau+galindo-2008,kero_mu_catalogue,schult_2018_meteoroid}.

Long-running southern-hemisphere surveys are essential because the viewing geometry and sporadic-source visibility depend strongly on latitude. The Advanced Meteor Orbit Radar and the Southern Argentina Agile Meteor Radar have produced important southern radiant and orbit surveys using the multi-static specular meteor trail echo method \citep{galligan2005radiant,janches2013initial,janches2015southern}, but high-power head-echo observations have generally been conducted as campaigns \citep{chau2007jicamarca_sources, kastinen_pansy} or been limited to several hundreds of events \citep{janches2014saamer_head,michell2015saamer_head,panka2021saamer_head}. The PANSY radar is located at Syowa Station, Antarctica, at approximately $69^\circ$ S. It therefore observes the southern apex and toroidal sources at favorable elevations and samples radiants that are poorly represented in northern head-echo catalogues.

This article describes an ongoing catalogue collection effort  rather than a single-event experiment. Its principal products are: 1) a continuously growing archive of voltage cuts that permits the complete reanalysis of every event, 2) pulse-level observables and fitted positions, and 3) event-level radiants, speeds, orbits, peak SNRs, and uncertainties. We describe the measurement and reduction methods in Section~\ref{sec:methods}, present the catalogue and selected compact stream structures in Section~\ref{sec:results}, interpret the radiant and initial-height morphology, selection effects, relative source rates, shower associations, and prospects for improved Doppler and dynamic-mass constraints in Section~\ref{sec:discussion}, and summarize the principal results in Section~\ref{sec:conclusions}.

\section{Observations and Methods}
\label{sec:methods}

PANSY is a 47~MHz mesosphere--stratosphere--troposphere and incoherent-scatter radar at Syowa Station \citep{sato2014program}. In January 2025, we installed an eight-channel receiver that provides the necessary interferometric angle of arrival observing capability needed for meteor observations. This allowed parallel observation of meteors without disrupting routine atmospheric observations.

 The standard PANSY operations interleave two-observing modes: a) upper-troposphere and lower-stratosphere (UTLS), and b) mesosphere. Of these two modes, the mesospheric mode is more suitable for meteor head echo observations, given its 1.6 ms interpulse period (IPP). The principal parameters used for meteor observations are summarized in Table~\ref{tab:parameters}. The catalogue presented here has used this mode. The experiment cycles through zenith, north, east, south, and west pointings with the transmit direction changing every IPP. The four off-zenith beams are displaced by $10^\circ$, and a given pointing is therefore revisited every 8~ms. Each transmitted pulse is a 128~$\mu$s 16-bit complementary coded waveform with 8 $\mu$s bits.

The mesospheric mode obtains about 9~h of observing time per day. A typical mesospheric mode window lasts 86~s, followed by 132~s of UTLS observations. This gives a start-to-start cycle of about 218~s and a mesospheric duty cycle of about 39\%. Approximately once a month, the radar runs a 100\% duty-cycle ionospheric observation for approximately 24 hours \citep{hashimoto}, which has not been analyzed for meteor head echoes.

\begin{figure}
\centering
\includegraphics[width=\linewidth,trim=34bp 12bp 72bp 10bp,clip]{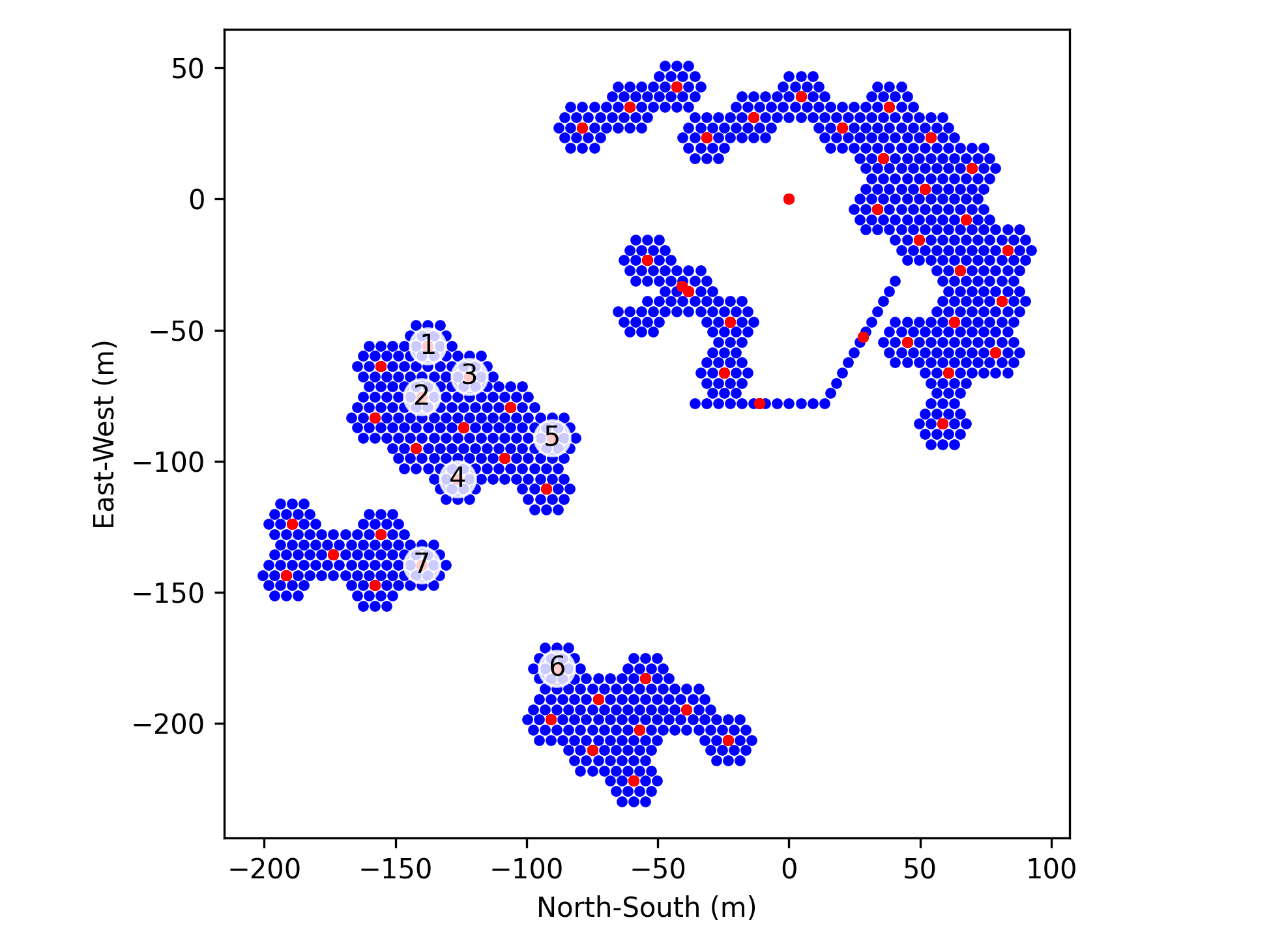}
\caption{PANSY antenna layout. The seven independently sampled receiver modules form the meteor interferometer.}
\label{fig:antenna_array}
\end{figure}

\begin{figure*}
\centering
\meteorpositionrow{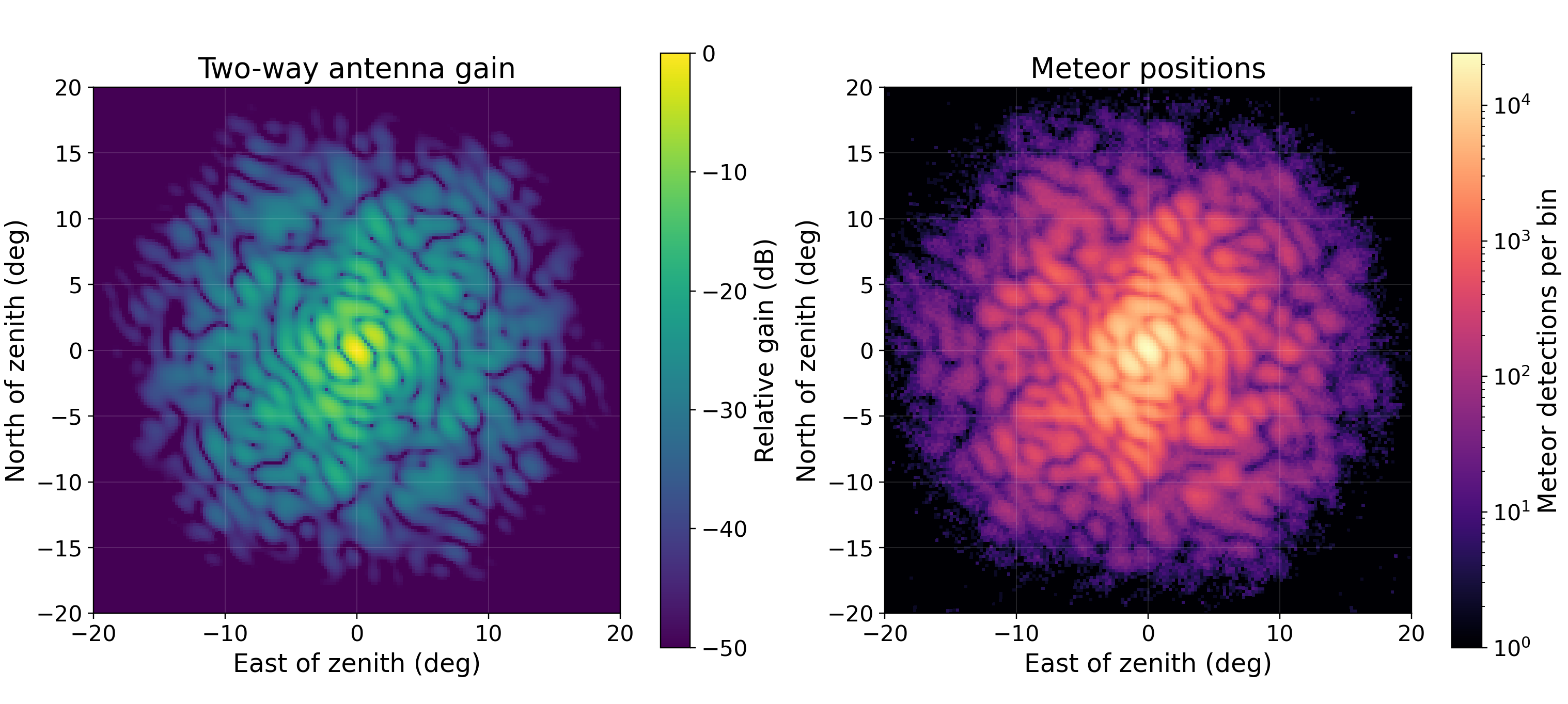}{a)}{b)}
\meteorpositionrow{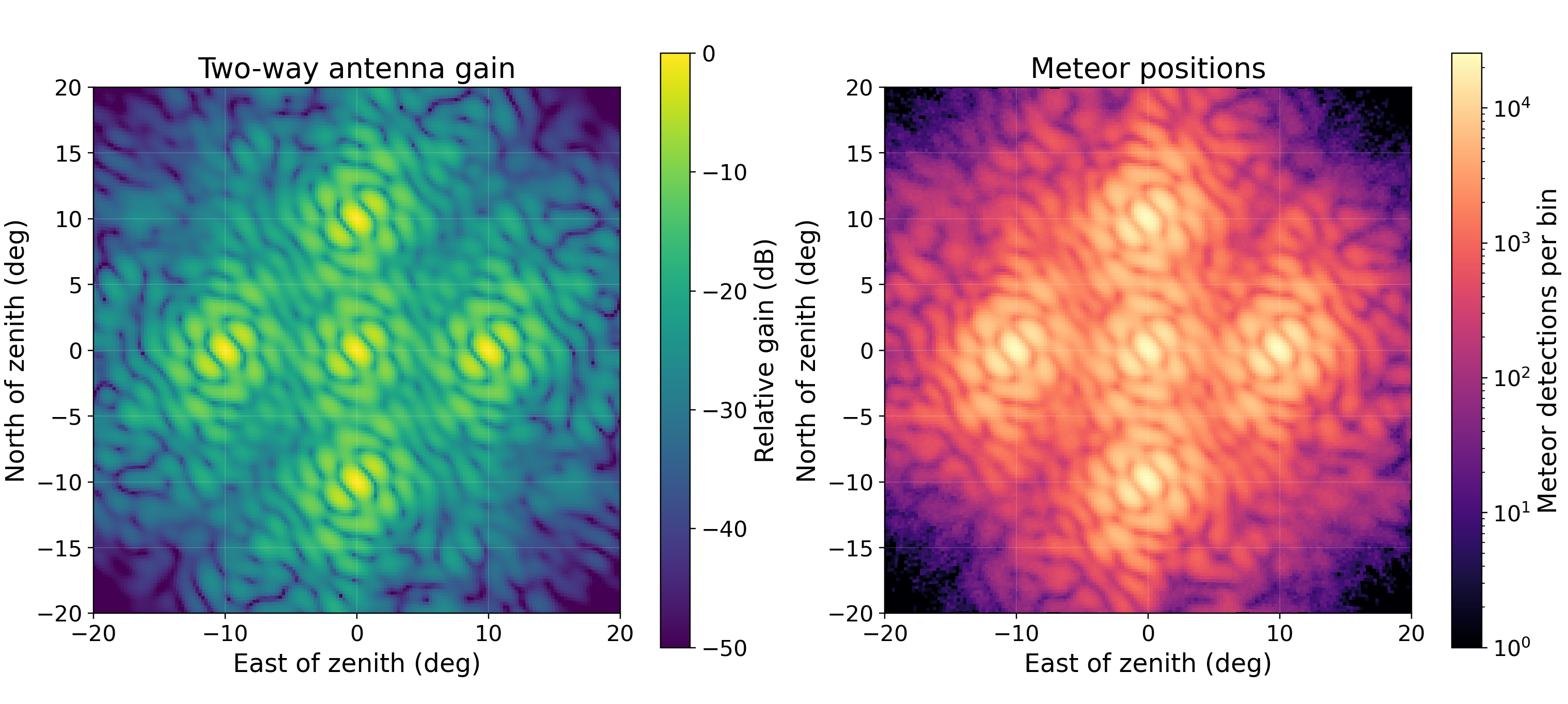}{c)}{d)}
\caption{Modeled two-way antenna gain and fitted meteor head-echo trajectory positions on the direction-cosine grid used for the interferometric search. Panels a) and b) compare the zenith-beam gain with positions measured during zenith transmission; panels c) and d) compare the composite gain of all five transmit beams with positions measured using all five beams. The field of view extends $20^\circ$ from zenith in the east--west and north--south directions.}
\label{fig:meteor_positions}
\end{figure*}

The full transmitting array contains 55 modules and 1045 Yagi antennas. Seven modules consisting of 19 Yagi antennas are digitized independently for meteor interferometry, as shown in Figure~\ref{fig:antenna_array}; digitizing all 55 modules on receive would have provided better interferometric capability, but would have exceeded the available data-rate, real-time processing, and power budgets for continuous meteor operation. One additional channel digitizes a sample of the transmitted waveform. The transmit-reference channel provides pulse time, transmit waveform sample, radar-mode identification, and a phase reference. The radar has a sparse array due to the challenging drifting snow. The transmit array geometry has substantial sidelobes \citep{nishimura}. Those lobes enlarge the illuminated region and often produce long head-echo arcs, which are beneficial for estimating the radiant angle, but also make radar cross-section more difficult, due to frequency diffraction pattern sidelobe occurrence of echoes. Figure \ref{fig:meteor_positions} shows the analytical two-way transmit (all modules) -- receive (one module) diffraction pattern, as well as a histogram of observed meteor head locations, which shows that meteors are detected in a $\approx30^{\circ}$ wide angular region around the transmit beam center, with meteor positions tracing the trasmit-receive diffraction pattern.

\begin{deluxetable*}{lccc}
\tablecaption{PANSY Meteor-mode Parameters\label{tab:parameters}}
\tablewidth{\textwidth}
\tablehead{\colhead{Parameter} & \colhead{Value} & \colhead{Unit} & \colhead{Role in catalogue}}
\startdata
Center frequency & 47 & MHz & wavelength and Doppler conversion \\
Peak transmit power & 500 & kW & sensitivity \\
Transmit gain & 37 & dB & sensitivity and beam model \\
Meteor receive gain & 20 & dB & one 19-element module \\
Transmit pointings & 5 & --- & zenith and four at $10^\circ$ \\
System-noise temperature & 4550--21700 & K & time-dependent threshold \\
Detection threshold & 7.0 (8.5) & linear (dB) & matched-filter SNR \\
Coded-pulse duration & 128 & $\mu$s & single-pulse integration \\
Complex sample rate & 1 & MHz & 149.9 m native range spacing \\
Nominal duty cycle & 8 & \% & observing exposure \\
Minimum RCS at 100 km & $6.7\times10^{-5}$ ($-41.7$) & m$^2$ (dBsm) & nominal single-pulse limit\\
\hline
\enddata
\end{deluxetable*}

\subsection{Range--Doppler matched filtering and event formation}
\label{sec:matched_filter}

Processing begins by identifying strong, compact echoes within individual transmitted pulses as meteor head-echo candidates. Each receiver channel is correlated with the measured transmit waveform over a grid of trial range delays and Doppler shifts. For receiver channel $c$, pulse time $t$, round-trip group delay $\tau$, and Doppler frequency $f$, the matched-filter sample is
\begin{equation}
 A_c(\tau,f;t)=
 \sum_{n=0}^{N-1}x_c(t+\tau+n\Delta t)s^*[n]
 e^{-i2\pi f n\Delta t}.
 \label{eq:matched_filter}
\end{equation}
Here $x_c$ is the complex voltage, $s$ is the measured transmitted code, and $\Delta t$ is the sample spacing. The seven receiver powers are summed incoherently,
\begin{equation}
 P(\tau,f;t)=\sum_{c=1}^{7}|A_c(\tau,f;t)|^2 .
 \label{eq:mf_power}
\end{equation}
This forms a range--Doppler search grid without first assuming a direction of arrival. The maximum of  Equation \ref{eq:mf_power} gives the pulse range and line-of-sight speed for each transmit pulse (indexed with $k$),
\begin{equation}
 R_k=\frac{c\hat\tau_k}{2}, \qquad
 v_{r,k}=\frac{c\hat f_k}{2f_0},
 \label{eq:range_doppler_observables}
\end{equation}
where $f_0$ is the radar frequency. The noise power $P_N$ is estimated independently for each pulse from ranges with no strong echoes, allowing a signal-to-noise ratio to be assigned to each echo:
\begin{equation}
 \mathrm{SNR}_k=\frac{P(\hat\tau_k,\hat f_k;t_k)-P_N}{P_N}.
 \label{eq:snr}
\end{equation}
Meteor events are formed by linking pulse detections whose ranges and Doppler shifts evolve smoothly, as expected for a meteor head echo. An event is accepted when it exceeds an 8.5~dB SNR threshold, contains at least ten thresholded pulse detections, and has candidate positions below 180~km altitude. For each accepted event, the seven receiver voltages and the transmit-reference voltage are copied from the raw voltage ring buffer with pre- and post-event padding. These voltage cuts are the temporary level~1 analysis product and preserve the original information needed for reprocessing while they remain available.

\subsection{Interferometric angle of arrival}
\label{sec:aoa}

The range--Doppler matched filter peak gives the target distance and radial speed, but not its position on the sky. The seven independently sampled receiver modules provide this angular information through their relative phases. At the selected range--Doppler peak, the calibrated phase for receiver pair $i,j$ is represented by the unit phasor
\begin{equation}
 z_{ij}=\frac{A_iA_j^*}{|A_iA_j^*|}=\exp(i\phi_{ij}).
 \label{eq:cross_phase}
\end{equation}
Let $\boldsymbol{r}_i$ denote the surveyed phase-center position of receiver module $i$ in the local Cartesian array frame. For a trial line-of-sight direction $\hat{\boldsymbol{s}}=(u,v,w)$ expressed in the same frame, the baseline is $\Delta\boldsymbol{r}_{ij}=\boldsymbol{r}_j-\boldsymbol{r}_i$, and we evaluate the phase coherence
\begin{equation}
 C(\hat{\boldsymbol{s}})=\left|\frac{1}{N_b}
 \sum_{i<j} z_{ij}\exp\left[-i\frac{2\pi}{\lambda}
 \Delta\boldsymbol{r}_{ij}\boldsymbol{\cdot}\hat{\boldsymbol{s}}\right]\right|.
 \label{eq:aoa}
\end{equation}
Here $N_b=21$ is the number of distinct receiver-module baselines in the sum. The search is carried out over directions above the local horizon. For pulse $k$, a coherence maximum defines the line-of-sight direction $\hat{\boldsymbol{s}}_k$, which is combined with the matched-filter range to give the Cartesian position
\begin{equation}
 \boldsymbol{x}_k=R_k\hat{\boldsymbol{s}}_k.
 \label{eq:position}
\end{equation}
The sparse receiver geometry produces grating aliases: a single pulse can have several high-coherence directions. We therefore do not treat the largest single-pulse maximum as automatically correct. Instead, all plausible angle-of-arrival maxima are carried forward as candidate directions and the ambiguity is resolved using the time continuity of the complete meteor trajectory, including its consistency with a straight atmospheric path and the final trajectory fit \citep{kastinen2020probabilistic}.

Figure~\ref{fig:meteor_positions} is therefore an important quality-control diagnostic for the interferometric solution. The histogram of estimated meteor positions form the same main-lobe and sidelobe structure as the analytic two-way transmit beam pattern, indicating that the event-level interferometric selection is successful for most meteors. The upper panels isolate zenith-beam transmissions; in these data the meteor distribution follows the modeled zenith beam closely, with clear divergence only beyond about $15^\circ$ from the beam axis. At such large off-axis angles, the ideal analytic diffraction pattern is not necessarily expected to be a faithful representation of the true installed transmit array. Additionally, only a very small fraction of meteor positions occur that angles larger than $15^\circ$ off the transmit beam axis.

Static intermodule phase corrections are determined separately for each transmit beam using a subset of meteor echoes. The calibration jointly maximizes the mean interferometric coherence and minimizes the difference between the observed mean meteor location and the expected mean location at the transmit-beam center, following an approach similar to the empirical meteor-echo phase calibration of \citet{chau_2019_empirical}.

\subsection{Trajectory fitting and alias selection}
\label{sec:trajectory}

Local maxima of Equation~\ref{eq:aoa} are linked into time-continuous path hypotheses. A preliminary robust fit rejects interferometric trajecotry hypotheses that cross the local horizon, with curved trajectories, or are in upward motion. Among the remaining candidates, the alias with the smallest reduced residual statistic is selected.

The selected path is then fitted with a single-body, shrinking-radius model of meteor drag and ablation, following the standard formulation reviewed by \citet{ceplecha1998meteor}. For meteoroid radius $r$, bulk density $\rho_m$, atmospheric density $\rho_a$, and apparent ablation coefficient $\sigma$, the model can be written
\begin{align}
 m(t)&=\frac{4\pi}{3}\rho_m r(t)^3,\label{eq:mass_radius}\\
 \frac{d\boldsymbol{v}}{dt}&=-
 \frac{3\rho_a}{4\rho_m r}|\boldsymbol{v}|\boldsymbol{v},\label{eq:ceplecha_drag}\\
 \frac{dm}{dt}&=-\rho_a\pi r^2\sigma|\boldsymbol{v}|^3.
 \label{eq:ceplecha_ablation}
\end{align}
These are the spherical reductions of Equations~(1) and~(2) of \citet{ceplecha1998meteor}, with free-molecular drag coefficient $\Gamma=1$ and apparent ablation coefficient $\sigma=\Lambda/(2\xi\Gamma)$. The atmospheric density is evaluated with NRLMSISE-00 \citep{picone2002nrlmsise}. We adopt fixed $\rho_m=3000~\mathrm{kg\,m^{-3}}$ and $\sigma=10^{-8}~\mathrm{kg\,J^{-1}}$; the latter is close to the apparent ablation coefficient tabulated by \citet{ceplecha1998meteor}. These material constants set the dynamic-mass scale and are not estimated separately for each meteor.

\begin{figure*}
\centering
\includegraphics[width=\linewidth]{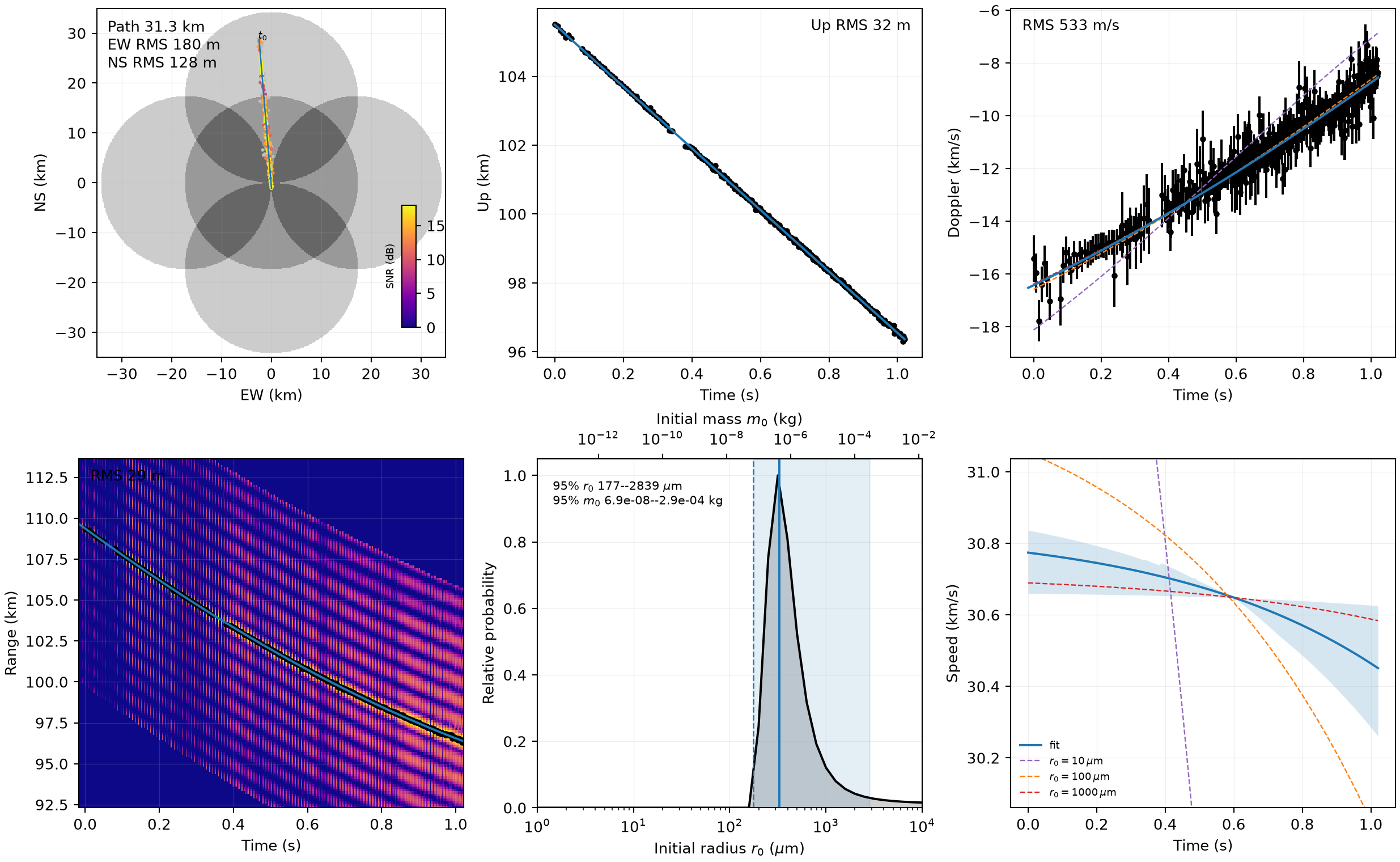}
\caption{Representative head echo fitted with the shrinking-radius drag--ablation model of Equations~\ref{eq:mass_radius}--\ref{eq:ceplecha_ablation}. The panels show the horizontal path and transmit-beam footprints, altitude, independently fitted single-pulse Doppler measurements, range--time matched-filter power, the marginal probability density of initial radius (with equivalent initial mass for $\rho_m=3000$~kg~m$^{-3}$), and speed. The solid blue curve is the selected shrinking-radius trajectory. Dashed curves are profile fits in which the initial radius $r_0$ is held at the indicated value; radius still evolves through ablation in every curve. The event excludes a very small initial radius, while retaining a broad range of larger radii and masses.}
\label{fig:example_detection}
\end{figure*}

Figure~\ref{fig:example_detection} shows an example 31.3~km long head-echo path observed on September 9th 2025. The shrinking-radius trajectory is constrained jointly by interferometric positions, range, and the independent single-pulse Doppler measurements. The fitted position residuals are 180, 130, and 32~m in the east--west, north--south, and vertical components, respectively, and the Doppler RMS residual is 530~m~s$^{-1}$. The event illustrates that the dynamics constrain the lower limit of the meteoroid radius more strongly than the upper limit. The $r_0=10\,\microm$ hypothesis predicts far more deceleration than observed and is excluded, while the $r_0=100\,\microm$ and $r_0=1000\,\microm$ hypotheses approach the measured evolution. Profiling the $\chi^2$ with a fixed $r_0$ and fitting the remaining initial-state parameters gives an estimate of the 95\% confidence interval $r_0=180$--$2800\,\microm$, corresponding to $m_0=7\times10^{-8}$--$3\times10^{-4}$~kg for the adopted density. The broad upper tail makes clear that the initial dynamic mass is only weakly constrained, as the majority of the radial acceleration is not due to atmospheric drag, but due to the geometric path of the meteoroid, as discussed by \cite{kero2008determination}.

\subsection{Meteoroid orbit}
\label{sec:orbit}

The selected atmospheric trajectory is fitted in the local radar frame. Its speed at the first retained pulse is denoted $v_0=|\boldsymbol{v}(t_0)|$; this local initial-detection speed is used for the atmospheric height--speed distribution in Figure~\ref{fig:height_velocity}. For radiant and orbit calculation, the fitted state at $t_0$ is transformed to a Cartesian Geocentric Celestial Reference System (GCRS) state, including the velocity contribution from Earth's rotation. Numerical propagation with REBOUND/IAS15 \citep{rein_liu_2012_rebound,rein_spiegel_2015_ias15} to the Sun-Earth Hill sphere distance is then used to remove Earth's gravitational acceleration (zenithal attraction) and obtain the asymptotic geocentric velocity $\boldsymbol{v}_g$. The geocentric radiant points opposite to $\boldsymbol{v}_g$. If $(\lambda_g,\beta_g)$ is the radiant in ecliptic coordinates and $\lambda_\odot$ the solar longitude, the sun-centered longitude coordinate is
\begin{equation}
 \lambda'_g=(\lambda_g-\lambda_\odot)\bmod 360^\circ,
 \qquad \beta'_g=\beta_g.
 \label{eq:sun_centered}
\end{equation}
The numerical propagation to the Sun--Earth Hill-sphere distance yields the asymptotic state used to determine the meteoroid's heliocentric orbit. At the propagation endpoint, the state is transformed to the heliocentric mean-ecliptic frame using the planetary ephemeris, and the osculating elements $a,e,i,\Omega,\omega$, true anomaly $\nu$, and perihelion distance $q$ are calculated. The uncertainties of the Keplerian parameters are estimated from ten propagated state realizations per meteor: the nominal Cartesian initial state at atmospheric entry and nine random vectors drawn from its estimated covariance matrix. Each realization undergoes the same numerical propagation used to remove zenithal attraction.

\subsection{Catalogue levels}
\label{sec:data_levels}

The analysis products are event based and divided into the levels listed in Table~\ref{tab:data_levels}. Level~1 is a temporary raw-voltage cut record that retains transmit waveforms and radar echoes for detected meteor events and allows reprocessing without loss of information while it remains available. Level~2 contains direct pulse observables: position and Doppler shift, and SNR as a function pulse time. Level~3 is the compact orbital-elements catalogue used for orbit and population studies. Every level carries the event timestamp, allowing unique identification.

Level~3 stores the covariance of the local trajectory parameters, including initial position and velocity, as well as the covariance matrix of the initial Cartesian state vector. It also stores standard deviations and a covariance matrix for the Keplerian parameters.

\begin{deluxetable*}{l l l}
\tablecaption{PANSY Meteor Catalogue Data Model\label{tab:data_levels}}
\tablehead{\colhead{Level} & \colhead{Product} & \colhead{Principal fields}}
\startdata
1 & Temporary complex-voltage cut & \parbox[t]{0.65\textwidth}{Seven receiver channels; transmit-reference channel; sample rate; pulse times; transmitted code; beam identifiers; event padding; calibration identifier.} \\
2 & Pulse measurement series & \parbox[t]{0.65\textwidth}{UTC and relative time; range and uncertainty; Doppler shift and uncertainty; linear and decibel SNR; noise estimate; ENU position and covariance; coherence; selected alias; quality flags.} \\
3 & Trajectory and orbit & \parbox[t]{0.65\textwidth}{Initial Cartesian GCRS state; local trajectory covariance; fit statistic; geocentric radiant $(\alpha_g,\delta_g)$; Sun-centered ecliptic radiant $(\lambda'_g,\beta_g)$; $v_g$; Keplerian elements $(a,e,i,\Omega,\omega,\nu,q)$; orbital-element standard deviations and covariance; observing geometry and quality flags.} \\
\hline
\enddata
\end{deluxetable*}

\subsection{Sensitivity and system noise}
\label{sec:sensitivity}

The radar detection sensitivity is set by the transmit pulse compression, range, antenna gain, and system-noise temperature. For a coded pulse of duration $T_p$, the nominal single-pulse minimum detectable radar cross section (RCS) follows from the matched-filter form of the monostatic radar equation,
\begin{equation}
 \sigma_{\min}=\frac{(4\pi)^3R^4 k_B T_s}{P_tG_tG_r\lambda^2T_p}
 \mathrm{SNR}_{\min},
 \label{eq:radar_sensitivity}
\end{equation}
where $T_s$ is system-noise temperature. With the nominal values in Table~\ref{tab:parameters}, and $R=100$~km, Equation~\ref{eq:radar_sensitivity} gives $6.7\times10^{-5}$~m$^2$, or $-41.7$~dBsm as the smallest detectable radar cross-section. In practice, due to relatively high gain antenna diffraction pattern sidelobes shown in Figure \ref{fig:meteor_positions}, a large fraction of the detections will be in the sidelobes, where the detection requires 10-20 dB larger radar cross-sections (-30 to -20~dBsm).

\begin{figure}
\centering
\includegraphics[width=\linewidth]{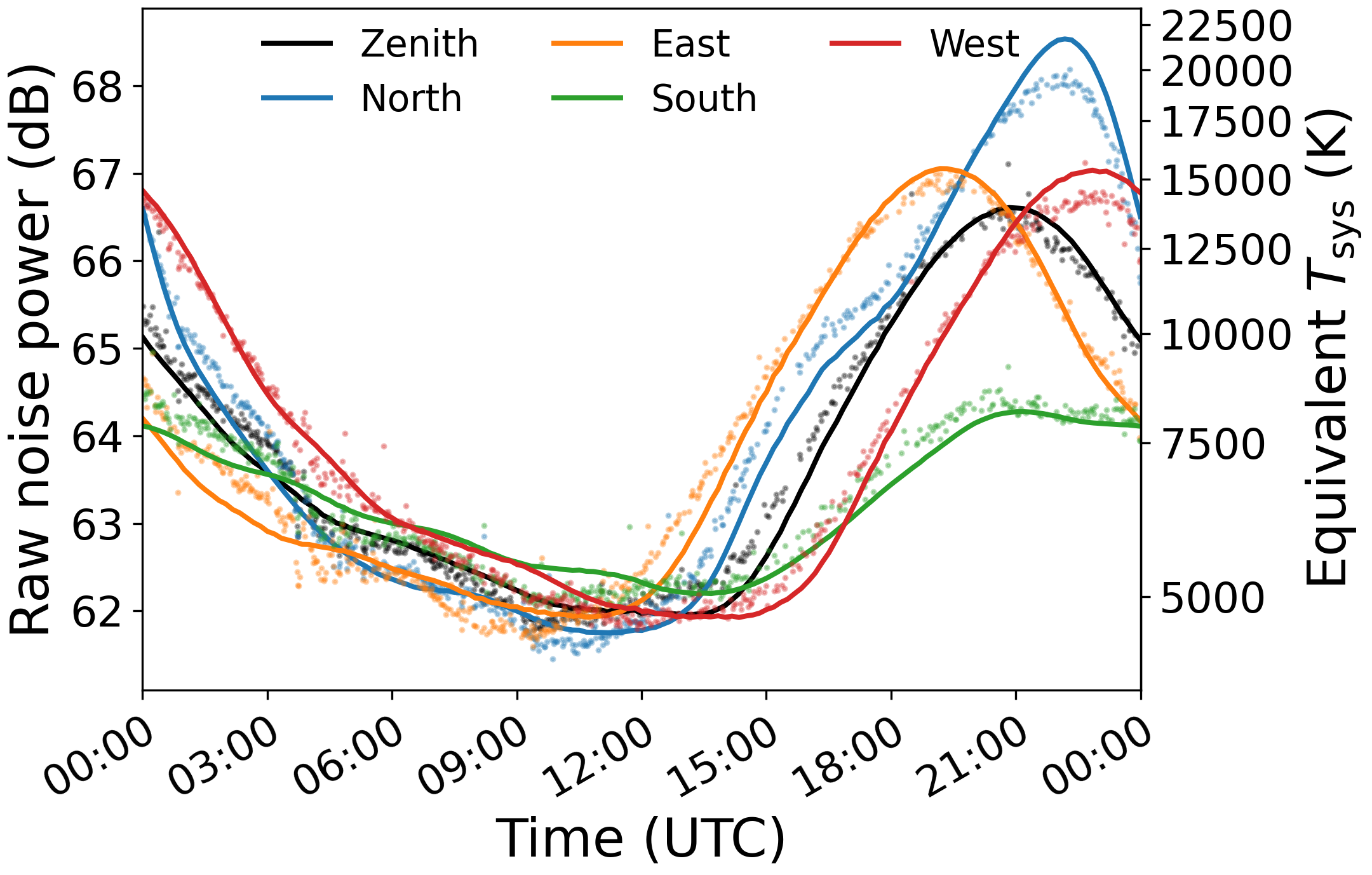}
\caption{Single module power measurements and model of system noise temperature for the five beam pointing directions.}
\label{fig:noise_model}
\end{figure}

The system-noise term is estimated from raw voltage samples taken at 1 MHz sample rate not containing meteor echoes and comparing the resulting power over a sidereal day with a beam-weighted sky-noise model. The sky contribution is computed from the Global Sky Model \citep{de2008model} convolved with the one-way receive pattern of a 19-antenna module, while the receiver temperature is fitted as an additive term. Figure~\ref{fig:noise_model} shows this calibration for one representative day. The system noise temperature is between 5000-22500 K, depending on the location of the galactic plane relative to the receiver module pointing direction. The variation of system noise also creates a sidereal day variability in detection sensitivity of 2-6.5 dB, depending on the transmit beam.

\section{Results}
\label{sec:results}

The database contains $\approx 2.1\cdot 10^6$ fitted meteor head-echo trajectories from the observing interval 2025 January 26--2026 July 26, comprising $\approx 54\cdot 10^6$ pulse-resolved position and radial-Doppler measurements. Unless otherwise stated, we have used 60\% of the best meteor events based on uncertainty and data quality. Each retained event has a valid winning-alias orbit solution that passes the trajectory-fit and physical-admissibility checks, an initial-state position standard deviation $\leq1000$~m, an initial-state radiant-angle standard deviation $\leq3^\circ$, and a first selected trajectory point within $20^\circ$ of the active transmit-beam center.

\begin{figure}
\centering
\includegraphics[width=\linewidth]{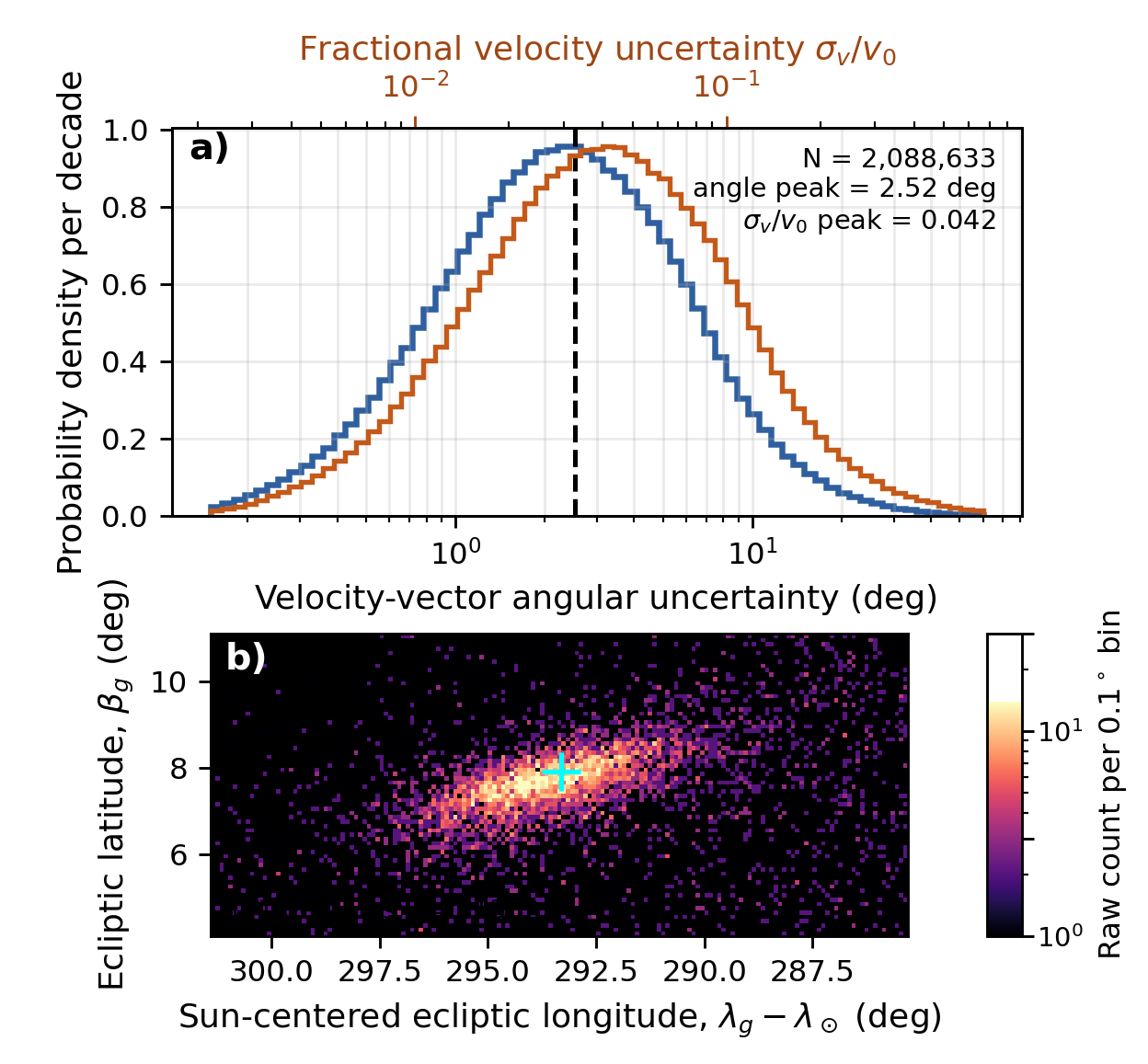}
\caption{a) Distribution of radiant angle and speed uncertainty for the full catalogue. b) Raw-count radiant histogram near the $\eta$-Aquariids, used to demonstrate radiant compactness, using all meteor radiants measured during $14^\circ\leq\lambda_\odot\leq100^\circ$ with 0.1~deg histogram bins. The cyan $+$ marker indicates radiant peak location from \citet{jenniskens_2023_atlas}.}
\figureprovenance{plot\_figure5\_uncertainty\_eta.py}
\label{fig:radiant_velocity_uncertainty}
\end{figure}

The radiant-angle uncertainty is a conservative velocity-vector estimate
\begin{equation}
\sigma_\theta = \tan^{-1}\left(\frac{\sigma_v}{v_0}\right),
\qquad
\sigma_v = \sqrt{\mathrm{tr}\!\left(\mathbf{C}_{vv}\right)},
\label{eq:radiant_angle_uncertainty}
\end{equation}
where $\mathbf{C}_{vv}$ is the local-velocity block of the fitted state covariance matrix and $v_0$ is the fitted local speed, both evaluated at the first retained pulse. Thus $\sigma_\theta$ describes uncertainty in the fitted atmospheric velocity vector angle at $t_0$. Figure~\ref{fig:radiant_velocity_uncertainty}a shows the angular uncertainty $\sigma_\theta$ on the lower axis and the fractional velocity uncertainty $\sigma_v/v_0$ on the upper axis. The $3^\circ$ quality threshold retains 61.6\% of the trajectory fits. To demonstrate angular resolution within the catalogue, Figure~\ref{fig:radiant_velocity_uncertainty}b shows the $\eta$ Aquarids meteor shower radiant histogram, without any exclusion of data in the catalogue, other than selecting the solar longitude interval that the shower is active ($14^{\circ} < \lambda_{\odot}< 100^{\circ}$). While much of the longitudinal extent of the shower is due to migration of the radiant, the ecliptic latitude extent is consistent with the estimated angular uncertainty.

\begin{figure}
\centering
\includegraphics[width=\linewidth]{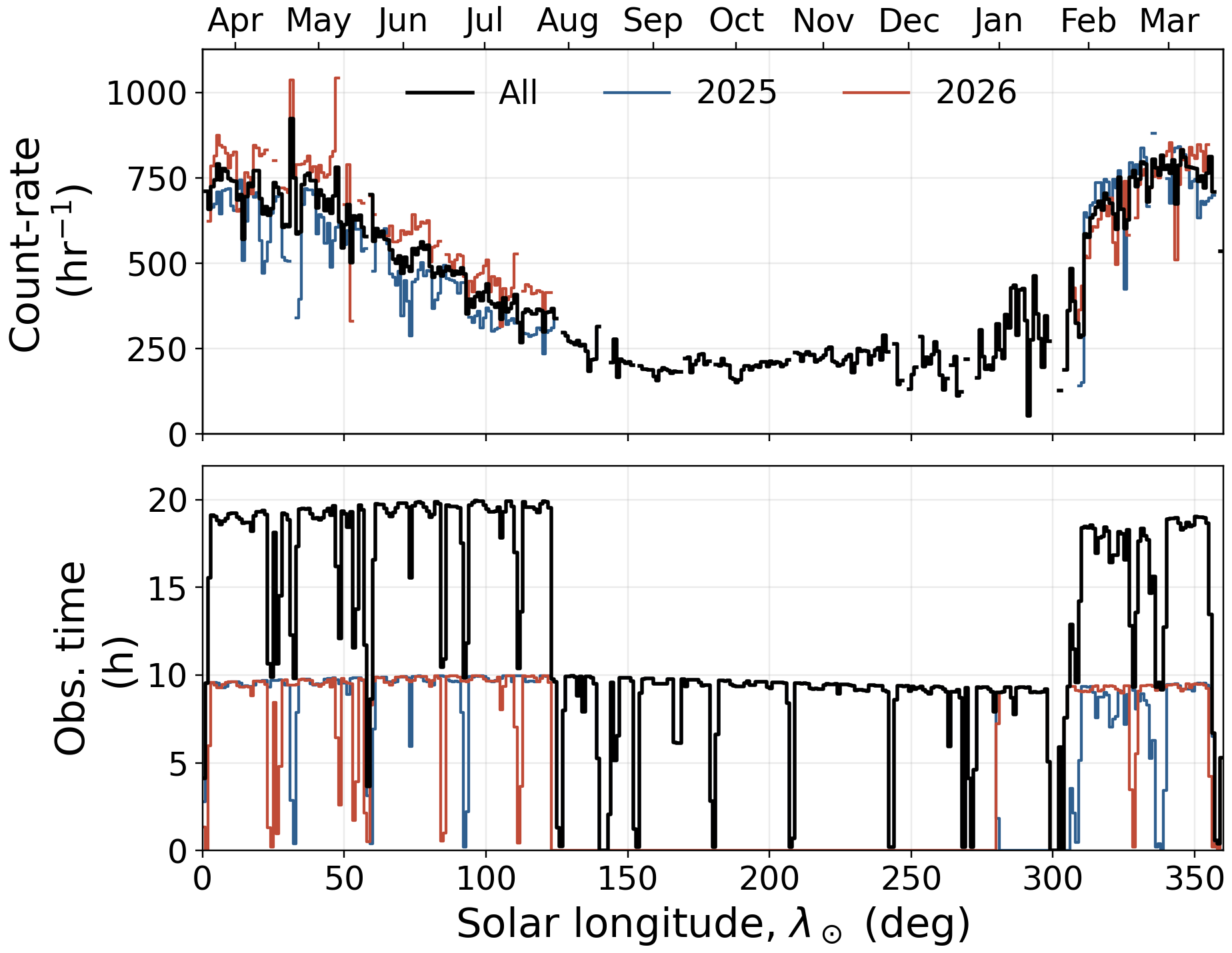}
\caption{Catalogue event rate and observing time as a function of solar longitude. The upper panel shows event counts per hour in $1^\circ$ bins for all observations and for the 2025 and 2026 subsets. Rates are shown only for bins containing at least 5~h of mesosphere-mode observations with completed orbit processing. The lower panel shows the actual mesosphere-mode observing time, including intervals awaiting orbit processing. The rate corrects for variations in observing duration but is not an absolute meteoroid flux.}
\figureprovenance{\textasciitilde/src/pansy\_receiver/plot\_orbit\_catalogue\_statistics.py}
\label{fig:counts_solar_longitude}
\end{figure}

Figure~\ref{fig:counts_solar_longitude} shows the catalogue event rate in $1^\circ$ bins of solar longitude. For each bin, the event count is divided by the duration of mesosphere-mode observations. The lower panel separately shows the actual mesosphere-mode observing time. The normalization removes the leading effect of observing gaps, but the resulting rate is not a calibrated flux: it remains modulated by sky noise, range and gain, viewing geometry, and the speed- and ablation-dependent detection probability. The rate has a broad maximum from February through May and a broad minimum from August through December.

\subsection{Detection height, speed, and radiant distribution}

\begin{figure}
\centering
\begingroup
\setlength{\unitlength}{\linewidth}
\begin{picture}(1,0.6944)
  \put(0,0){\includegraphics[width=\linewidth]{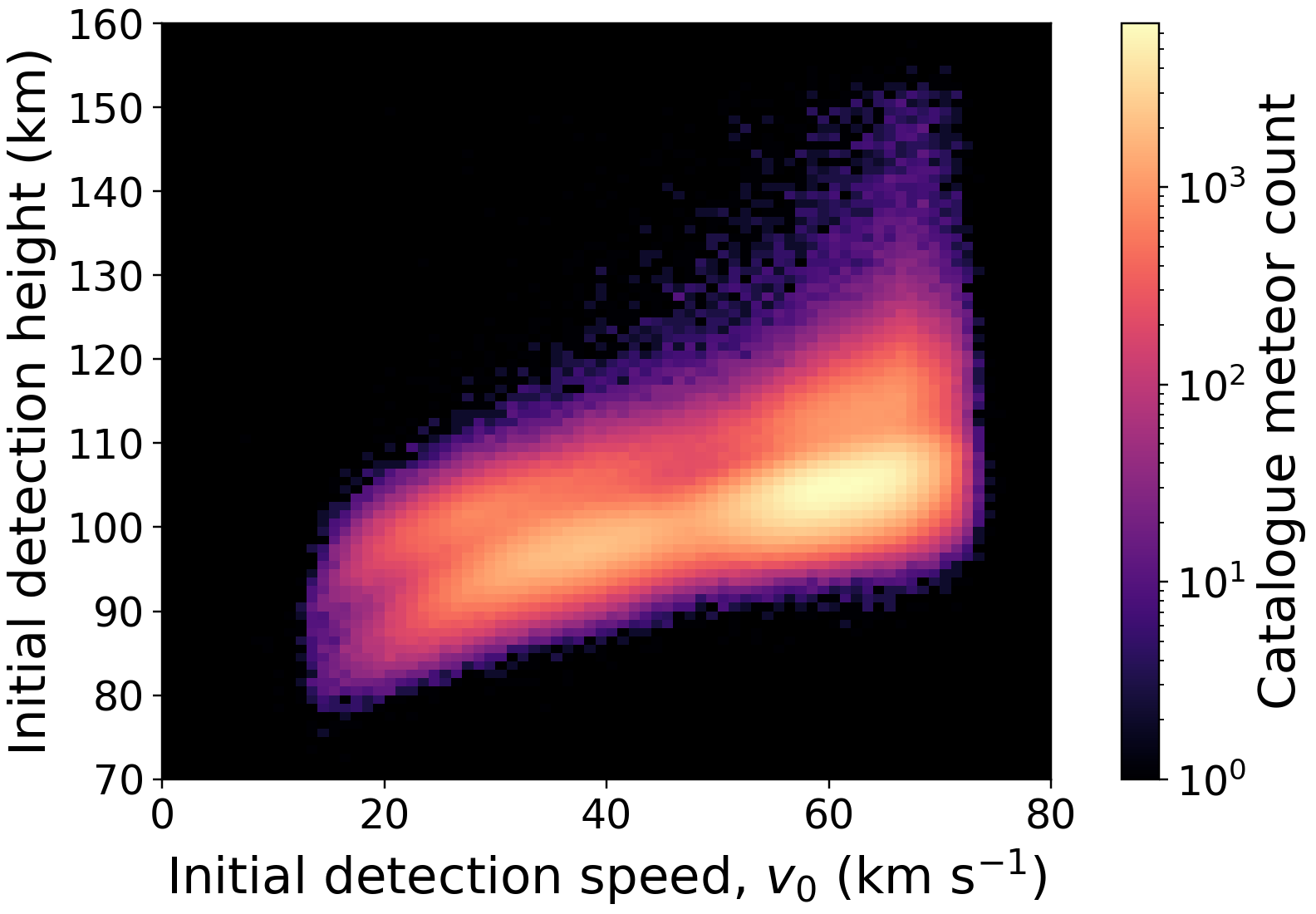}}
  \linethickness{0.8pt}
  \color{white}
  \put(0.165,0.486){\makebox(0,0)[l]{\textbf{Upper band}}}
  \Line(0.315,0.470)(0.376,0.326)
  \put(0.629,0.134){\makebox(0,0)[c]{\textbf{Lower band}}}
  \Line(0.625,0.150)(0.461,0.262)
\end{picture}
\endgroup
\caption{Initial detection height versus fitted local speed $v_0=|\boldsymbol{v}(t_0)|$ at the first retained pulse for quality-selected catalogue meteors. This is the atmospheric initial-detection speed, not the asymptotic geocentric speed $v_g$. Empty bins are black and populated-bin color is logarithmic in event count.}
\label{fig:height_velocity}
\end{figure}

The joint distribution of initial detection height and the fitted local speed $v_0$ at the first retained pulse is shown in Figure~\ref{fig:height_velocity}. The lower envelope rises with speed: slow meteors are commonly first detected below 100~km, whereas fast meteors are detected at higher altitude. The first-detection heights  form two distinct bands. Both bands are visible across the full speed range, but their relative prominence and separation vary with speed.

\begin{figure*}
\centering
\includegraphics[width=\linewidth]{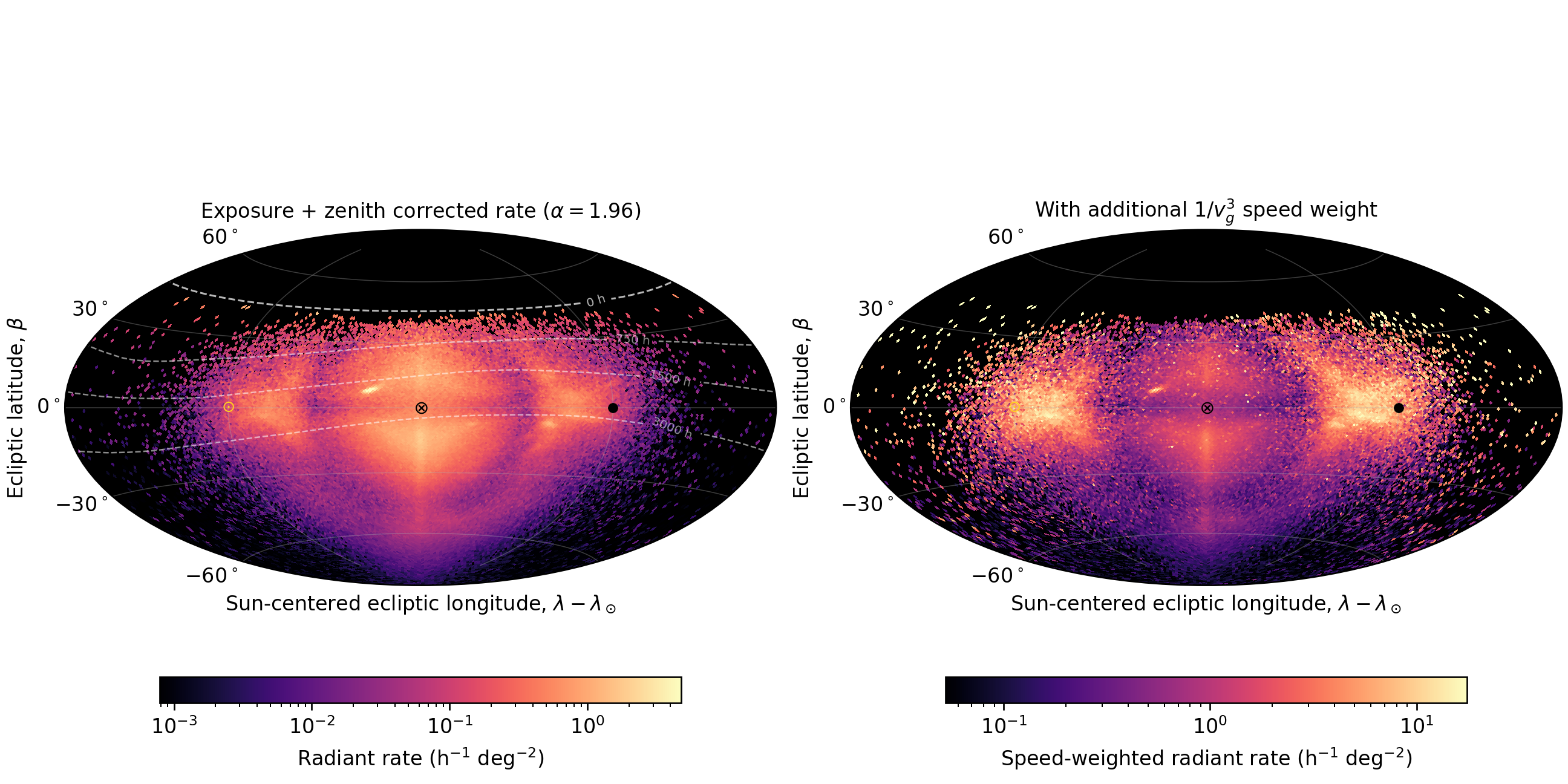}
\caption{Sun-centered ecliptic equal-angular-area radiant histogram with $0.839~\mathrm{deg}^2$ bins. Left: rate density after correcting for radiant exposure and zenith angle. Right: the same rate with an additional correction for the velocity dependence of radar sensitivity, which increases the inferred contribution of slow-moving meteoroids. The $\odot$, $\otimes$, and $\bullet$ mark the helion, apex, and antihelion directions, respectively.}
\figureprovenance{plot_paper_radiant_results.py}
\label{fig:radiant_distribution}
\end{figure*}

Figure~\ref{fig:radiant_distribution} displays $1.3\cdot 10^6$ high-quality radiants in Sun-centered ecliptic coordinates after two levels of empirical flux debiasing. All radiant maps use equal-angular-area histogram bins, implemented with HEALPix \citep{gorski2005healpix}; displayed counts and rates are divided by the angular bin area. The fine maps use bins of $0.839~\mathrm{deg}^2$, while the lower-resolution snapshots use bins of $3.36~\mathrm{deg}^2$. The left panel divides by radiant exposure time and assigns each event $i$ a zenith-angle weight
\begin{equation}
w_{z,i}=\left[\max(\cos z_{R,i},0.15)\right]^{-\alpha},
\label{eq:zenith_angle_weight}
\end{equation}
where $z_{R,i}$ is the local zenith angle of the meteor radiant at the detection epoch. This weight accounts for changes in the geometric projection of the incident flux onto the radar collection volume. We obtain $\alpha=1.96$ by minimizing the squared logarithmic difference between the exposure-corrected integrated northern and southern apex rates in mirrored regions with $|\lambda'_g-270^\circ|\leq30^\circ$ and $5^\circ\leq|\beta|\leq25^\circ$.

The right panel of Figure~\ref{fig:radiant_distribution} applies an additional velocity-dependent weight,
$w_{v,i}=(v_{\mathrm{ref}}/v_{0,i})^3$, where $v_{0,i}$ is the atmospheric entry speed of meteor $i$. The $v^{-3}$ form approximates the reduced detection sensitivity to slower meteors because the ablation rate and head-echo plasma density scale approximately as $v^3$ to first order. Here $v_{\mathrm{ref}}=73~\kms$, near the upper end of the velocity distribution, and the final bin values are
\begin{equation}
\Phi = \frac{1}{T_{\mathrm{rad}}}
\sum_i w_{v,i}
w_{z,i}.
\label{eq:radiant_rate_weight}
\end{equation}
This is only a first-order debiasing. A more complete speed correction would incorporate the radar cross section \citep{dimant2017formation} and the ablation rate from a physical ablation model \citep{vondrak2008chemical}. Such a correction was not tractable for the full catalogue yet because the initial radii are poorly constrained. The original raw radiant counts are shown in Figure~\ref{fig:sporadic_source_apertures}.

Both radiant maps shown in Figure~\ref{fig:radiant_distribution} contain the helion and antihelion concentrations and the apex complex. The broad southern coverage is the defining geometric contribution of this catalogue, as some southern ecliptic latitudes remain above the horizon for 24 hours per day. The radiant-bin exposure for the full observing interval is overlaid on the radiant distribution in hours. The exposure $T_{\mathrm{rad}}$ is the true radiant exposure, defined by the time that the radiant is above the horizon during the mesospheric radar mode.

\begin{figure*}
\centering
\begingroup
\setlength{\unitlength}{\linewidth}
\begin{picture}(1,0.6382)
  \put(0,0){\includegraphics[width=\linewidth]{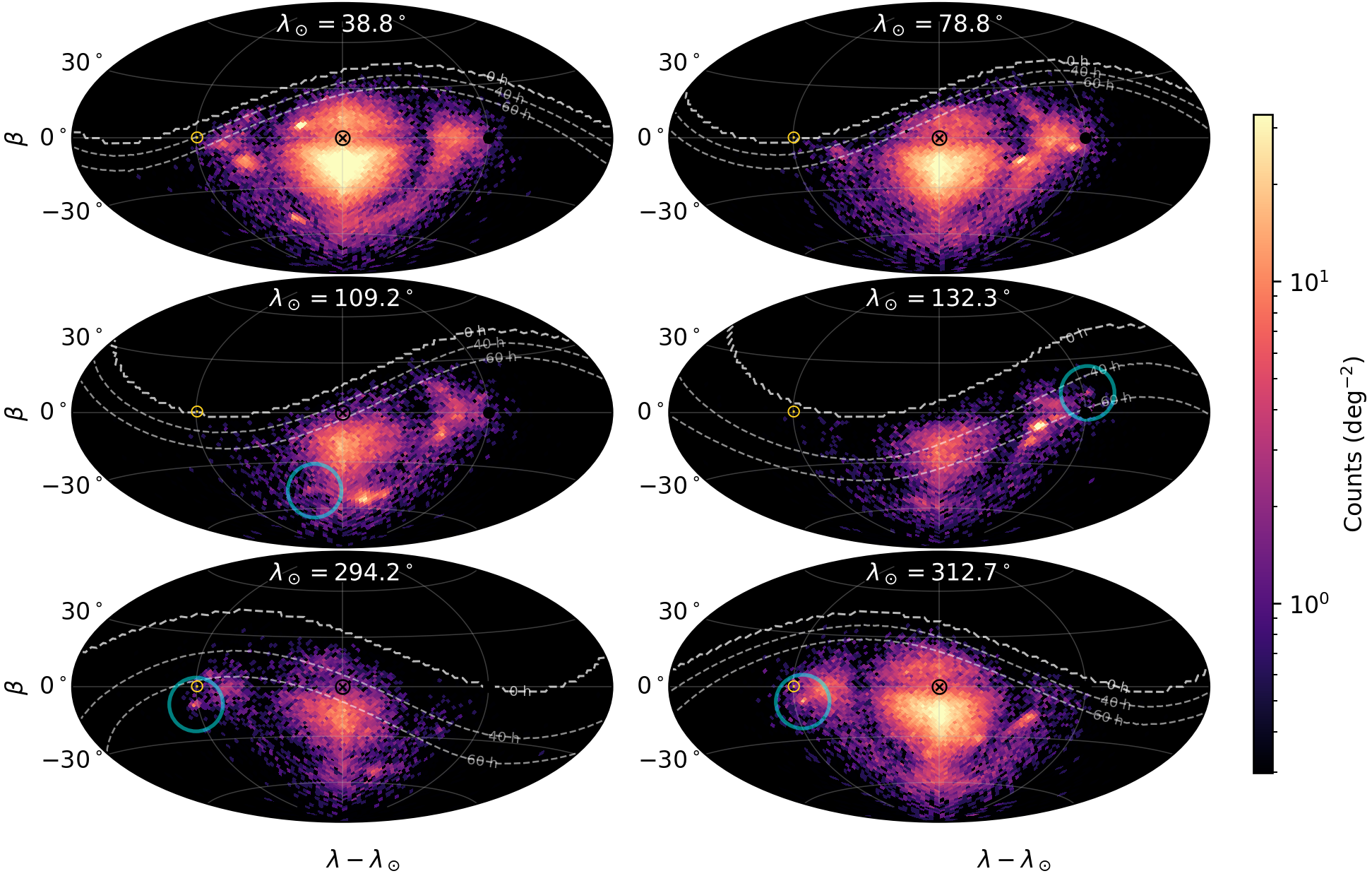}}
  \put(0.006,0.626){\makebox(0,0)[lt]{\textbf{a)}}}
  \put(0.475,0.626){\makebox(0,0)[lt]{\textbf{b)}}}
  \put(0.006,0.424){\makebox(0,0)[lt]{\textbf{c)}}}
  \put(0.475,0.424){\makebox(0,0)[lt]{\textbf{d)}}}
  \put(0.006,0.222){\makebox(0,0)[lt]{\textbf{e)}}}
  \put(0.475,0.222){\makebox(0,0)[lt]{\textbf{f)}}}
\end{picture}
\endgroup
\caption{Raw radiant-count snapshots at six representative solar longitudes, reported as counts per square degree using equal-area cells of area $3.36~\mathrm{deg}^2$. No exposure, zenith-angle, or velocity correction is applied. Radiant exposure is indicated with white contours. The panels show $\lambda_\odot=38.8^\circ$, $78.8^\circ$, $109.2^\circ$, $132.3^\circ$, $294.2^\circ$, and $312.7^\circ$, each using a $\pm4^\circ$ solar-longitude window. The $\odot$, $\otimes$, and $\bullet$ mark the helion, apex, and antihelion directions, respectively. Cyan circles mark the radiant regions of the July Fornax--Eridanus (JFE) shower candidate at $\lambda_{\odot}=109.2^{\circ}$, $\alpha$ Capricornids (CAP) at $\lambda_{\odot}=132.3^{\circ}$, and the extended Daytime Capricornids-Sagittariids (DCS) radiant at $\lambda_{\odot}=294.2^{\circ}$ and $312.7^{\circ}$.}
\figureprovenance{plot_paper_radiant_results.py}
\label{fig:radiant_snapshots}
\end{figure*}

\subsection{CAP and DCS adjacent to 169P/NEAT}
\label{sec:shower_results}

The $\alpha$ Capricornids (CAP) and Daytime Capricornids-Sagittariids (DCS; IAU shower 115) were selected to demonstrate the stream structure resolved by the catalogue. Both have well-defined radiants and orbital elements similar to comet 169P/NEAT, which has been proposed as their parent body \citep{brown_etal_2010_cmor_minor_showers,jenniskens_vaubaillon_2010_capricornids}.

CAP is visible in the $\lambda_\odot=132.3^\circ$ panel of Figure~\ref{fig:radiant_snapshots} near the Sun-centered ecliptic radiant $(\lambda'_g,\beta_g)=(177.2^\circ,10.8^\circ)$. The extended DCS radiant is visible in both daytime panels: near $(1.1^\circ,-9.8^\circ)$ at $\lambda_\odot=294.2^\circ$ and near $(354.5^\circ,-8.3^\circ)$ at $\lambda_\odot=312.7^\circ$.

\begin{figure*}
\centering
\includegraphics[width=\linewidth]{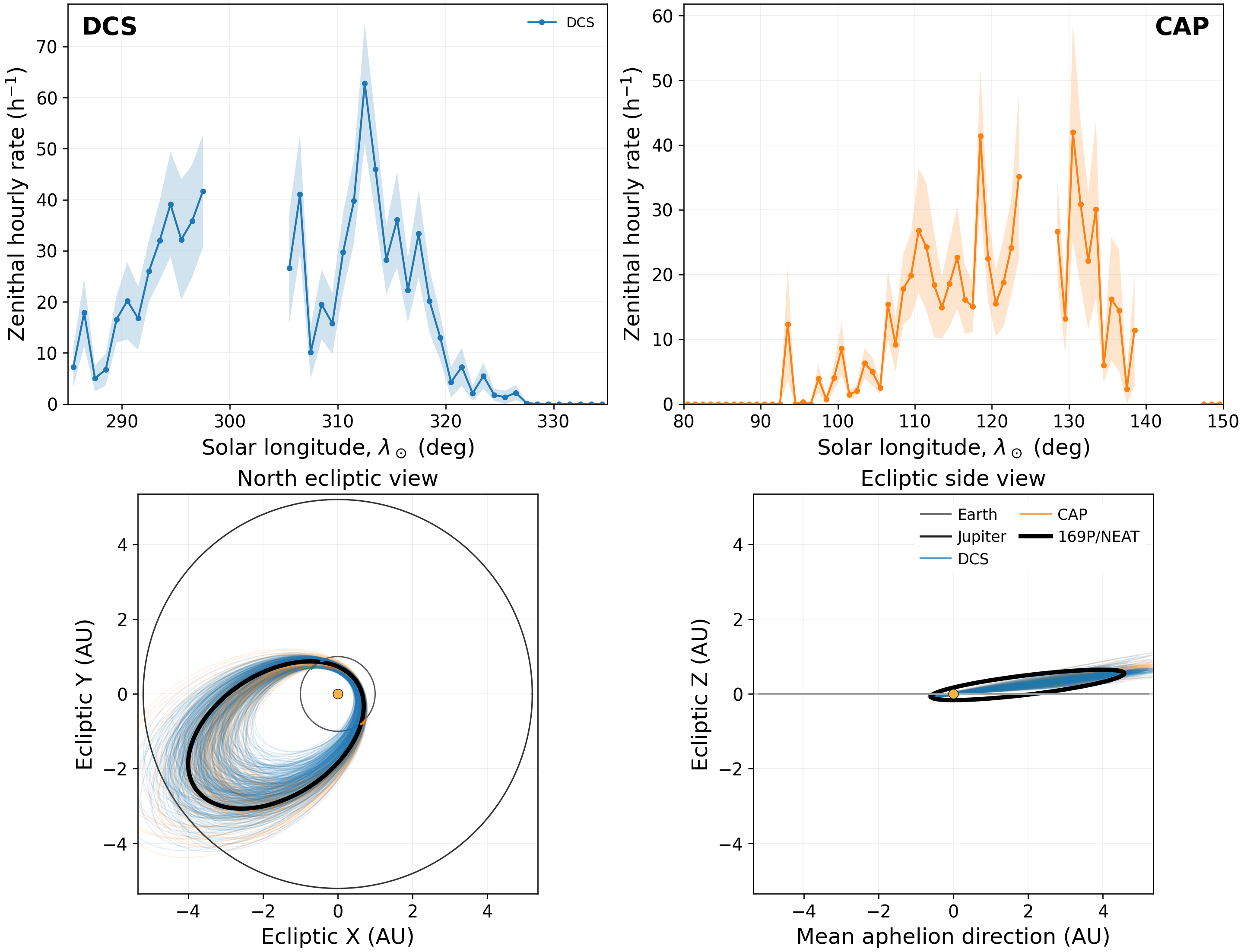}
\figureprovenance{plot_capricornid_conjugate_stream.py}
\caption{Activity profiles and measured orbits of the CAP--169P/NEAT complex. The top-left panel shows the extended DCS activity over $285^\circ\leq\lambda_\odot\leq335^\circ$ using one unchanged radiant--speed--orbit aperture. The top-right panel shows CAP activity over $80^\circ\leq\lambda_\odot\leq150^\circ$. Both curves are PANSY zenithal hourly rates; shaded regions give weighted Poisson standard errors, and intervals without sufficient radiant exposure are left blank. The bottom panels show the selected PANSY meteoroid orbits together with the orbit of 169P/NEAT, viewed from the north ecliptic pole and from the side with the horizontal axis aligned to the mean aphelion direction.}
\label{fig:capricornid_stream}
\end{figure*}

Figure~\ref{fig:capricornid_stream} presents the activity profiles and measured orbits of CAP and DCS. The upper panels give PANSY zenithal hourly detection rates in $1^\circ$ solar-longitude windows after applying the fitted zenith-angle weight and dividing by measured radiant exposure. One fixed radiant--speed--orbit aperture is used for DCS over the full $285^\circ$--$335^\circ$ interval: $v_g=18.5$--$28.5\,\kms$, $e=0.660$--0.845, $a=1.6$--3.6~AU, $i=4.0^\circ$--$10.5^\circ$, $q=0.40$--0.68~AU, $287^\circ\leq\alpha_g\leq323^\circ$, and $-36.5^\circ\leq\delta_g\leq-22.0^\circ$. PANSY accumulated 618.4~h of meteor-mode operation over this solar-longitude interval. Bins with less than three hours of radiant exposure, their immediately adjacent bins, and isolated valid runs shorter than three bins are omitted. The error band shows the Poisson counting uncertainty after applying the same zenith-angle weights used in the corrected rate. The lower panels show the corresponding PANSY meteoroid orbits and the reference orbit of comet 169P/NEAT.

The observed CAP enhancement spans approximately $106^\circ$--$139^\circ$ in solar longitude, while the extended DCS enhancement spans approximately $291^\circ$--$325^\circ$. DCS contains a localized enhancement near $313^\circ$, but incomplete exposure prevents a reliable measurement of the depth of the neighboring minimum. We therefore treat the daytime activity as one extended DCS radiant rather than dividing it into separate showers.

Table~\ref{tab:dcs_mean_parameters} lists the PANSY parameters and literature values for CAP and DCS. Angular quantities in Tables~\ref{tab:dcs_mean_parameters} and \ref{tab:ser_mean_parameters} are in degrees, $v_g$ is in $\kms$, and $a$ and $q$ are in AU; $\omega$ is the argument of perihelion and $\Omega$ is the longitude of the ascending node. The final column gives the Jopek \citeyearpar{jopek1993remarks} $D_H$ orbital dissimilarity from the listed 169P/NEAT reference orbit. The CAP rows are the \citet{brown_etal_2008_cmor_survey} and \citet{jenniskens_2023_atlas} values compiled in the IAU Meteor Data Center entry for shower 1. The DCS comparison rows are from \citet{brown_etal_2010_cmor_minor_showers} and \citet{jenniskens_2023_atlas}, as compiled in the IAU entry for shower 115. The PANSY DCS row characterizes the strongest localized portion of the extended radiant. The 169P/NEAT row contains the reference orbit used in Figure~\ref{fig:capricornid_stream}.

\begin{deluxetable*}{llrrrrrrrrrrrr}
\tabletypesize{\scriptsize}
\tablecaption{Comparison of meteor shower parameters with orbits adjacent to 169P/NEAT\label{tab:dcs_mean_parameters}}
\tablehead{
\colhead{Shower} &
\colhead{Source} &
\colhead{$\lambda_\odot$} &
\colhead{$\lambda_{\odot,\mathrm{act}}$} &
\colhead{$\alpha_g$} &
\colhead{$\delta_g$} &
\colhead{$v_g$} &
\colhead{$a$} &
\colhead{$q$} &
\colhead{$e$} &
\colhead{$\omega$} &
\colhead{$\Omega$} &
\colhead{$i$} &
\colhead{$D_H$}
}
\startdata
CAP & Brown et al. (2008) & 123.5 & \nodata & 302.9 & $-9.9$ & 22.2 & 2.35 & 0.59 & 0.750 & 269.2 & 123.3 & 7.3 & 0.159 \\
CAP & Jenniskens (2023) & 125.8 & \nodata & 304.8 & $-9.5$ & 22.7 & 2.74 & 0.58 & 0.790 & 267.8 & 126.0 & 7.7 & 0.153 \\
CAP & PANSY (this work) & 119.3 & 106--139 & 300.7 & $-10.6$ & 22.8 & 2.32 & 0.56 & 0.755 & 272.5 & 119.3 & 7.5 & 0.172 \\
\hline
DCS & Brown et al. (2010) & 301.0 & \nodata & 304.7 & $-29.2$ & 23.8 & 2.67 & 0.56 & 0.792 & 270.9 & 121.0 & 7.3 & 0.170 \\
DCS & Jenniskens (2023) & 311.5 & \nodata & 311.0 & $-26.9$ & 23.2 & 1.91 & 0.51 & 0.735 & 261.7 & 131.6 & 6.7 & 0.168 \\
DCS & PANSY (this work) & 313.6 & 291--325 & 312.4 & $-26.3$ & 25.3 & 2.29 & 0.48 & 0.790 & 259.9 & 133.6 & 7.5 & 0.180 \\
\hline
\nodata & 169P/NEAT reference orbit & \nodata & \nodata & \nodata & \nodata & \nodata & 2.60 & 0.60 & 0.768 & 218.1 & 176.0 & 11.3 & 0.000
\enddata
\end{deluxetable*}

\subsection{July Fornax--Eridanus candidate}
\label{sec:omega_eridanids}

The PANSY radiant maps contain several compact enhancements without an obvious counterpart in the IAU Meteor Data Center Working List. Here we examine the strongest of these candidates: a radiant straddling the formal Fornax--Eridanus constellation boundary near $\lambda_{\odot}=109.5^{\circ}$ and detected in both 2025 and 2026. Of the 138 selected radiants, 106 fall within Fornax and 32 within Eridanus according to the IAU constellation boundaries; the mean radiant lies in Fornax. Its main activity interval is $\lambda_\odot=108^\circ$--$112^\circ$. We therefore refer to the candidate as the July Fornax--Eridanus (JFE). The radiant is circled in Figure~\ref{fig:radiant_snapshots}, and its mean radiant, speed, Keplerian elements, and activity interval are listed in Table~\ref{tab:ser_mean_parameters}.

\begin{figure*}
\centering
\includegraphics[width=\linewidth]{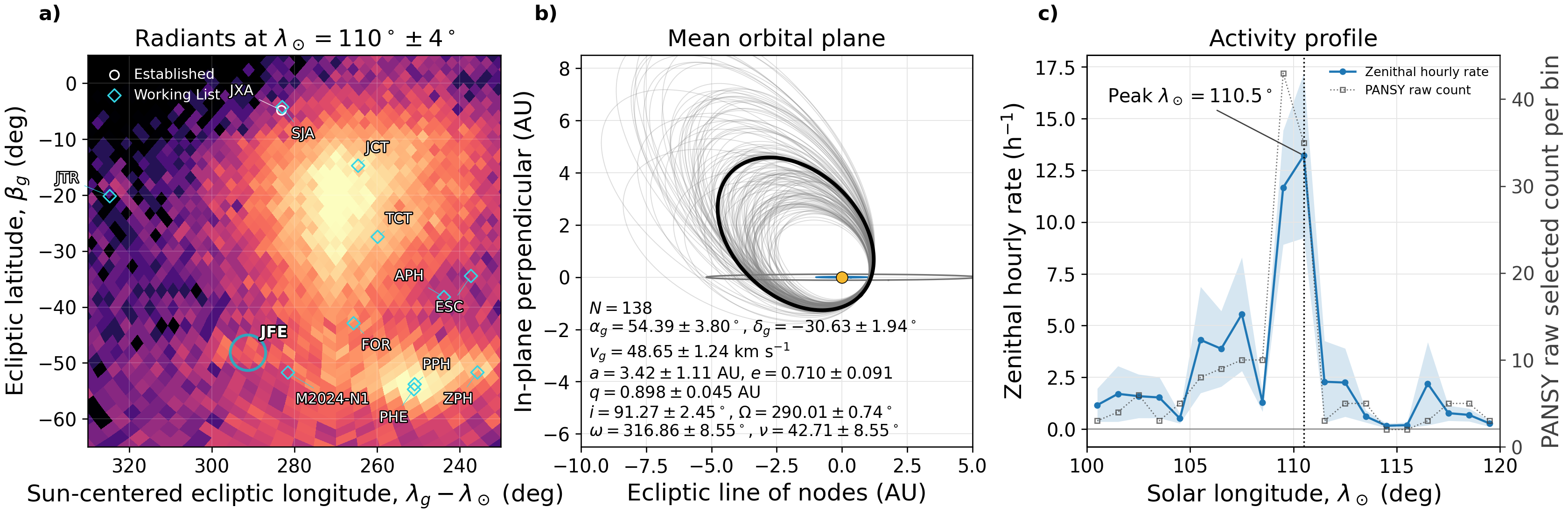}
\caption{Radiant distribution, orbit ensemble, and activity profile of the candidate July Fornax--Eridanus (JFE). Panel a shows the raw PANSY radiant-count density for $106^\circ\leq\lambda_\odot\leq114^\circ$ in Sun-centered ecliptic coordinates, with longitude displayed in the mirrored orientation used for the full radiant maps. The large cyan circle marks the JFE radiant region; small white circles and cyan diamonds mark nearby established and Working List shower radiants, respectively, whose catalogue peaks lie within $10^\circ$ of $\lambda_\odot=110^\circ$. Panel b projects the 138 selected JFE heliocentric orbits into the mean orbital plane, with the Sun at the focus and the intersection with the ecliptic plane horizontal; the thick curve is the mean orbit. The annotation gives sample means and standard deviations, using circular statistics for angular elements. In panel c, the blue curve gives the PANSY zenithal hourly rate in $1^\circ$ solar-longitude bins after correcting for radiant zenith angle and measured radiant exposure. The gray curve gives the corresponding raw PANSY count in the fixed radiant--speed--orbit selection.}
\figureprovenance{plot\_omega\_eridanids\_shower.py}
\label{fig:omega_eridanids}
\end{figure*}

The orbit panel of Figure~\ref{fig:omega_eridanids} shows a compact sample of 138 JFE meteors, comprising 75 detections in 2025 and 63 in 2026. The mean parameters in Table~\ref{tab:ser_mean_parameters} are calculated from this sample. The activity panel uses a broader fixed radiant--speed--orbit aperture so that the same selection can be evaluated both inside and outside the main shower interval. Each detection is weighted for radiant zenith angle, and the weighted count is divided by the measured radiant exposure, as in Figure~\ref{fig:capricornid_stream}. No background subtraction or speed-dependent sensitivity correction is applied.

We tested the activity enhancement against the local background using this broader fixed aperture, rather than the compact 138-meteor sample used for the orbit ensemble. The aperture contains 91 detections during the $108^\circ$--$112^\circ$ shower interval in 67.0~h of radiant exposure, giving $1.36\pm0.14~\mathrm{h}^{-1}$. The same aperture contains 48 detections during the preceding $100^\circ$--$108^\circ$ interval and 30 during the following $112^\circ$--$120^\circ$ interval. Their respective exposures are 154.1 and 150.9~h, giving rates of $0.31\pm0.05$ and $0.20\pm0.04~\mathrm{h}^{-1}$. Combining the adjacent intervals gives a local background rate of $0.26\pm0.03~\mathrm{h}^{-1}$ and approximately 17 expected background detections during the shower exposure. The observed count is therefore about five times the local-background expectation.

We compared JFE with the complete 2026 IAU Meteor Data Center Working List \citep{jenniskens2020workinglist}, updated 2026 August 10. The list contains 1,339 parameter sets. Repeated solutions for the same shower were averaged before comparison. Of those, 48 peak within $10^\circ$ in solar longitude of the JFE peak. These 48 entries were ranked by great-circle radiant separation from the JFE mean radiant. Table~\ref{tab:ser_mean_parameters} includes the four nearest radiants: M2024-N1, FOR, PPH, and PHE.

The two M2024-N1 solutions are from \citet{segon2024fornax} and \citet{roggemans2026m2024n1}; the PPH mean combines the solutions of \citet{pokorny_2017_saamer_stream_survey} and \citet{greaves2026meteor_candidates}, and the PHE mean combines those of \citet{cook1973workinglist} and \citet{jenniskens_2023_atlas}. The FOR parameters originate from an unpublished 2022 submission recorded in the Working List.

\begin{deluxetable*}{lrrrrrrrrrrrrr}
\tabletypesize{\scriptsize}
\tablecaption{July Fornax--Eridanus and Nearest Catalogue Comparisons\label{tab:ser_mean_parameters}}
\tablehead{
\colhead{Object or stream} &
\colhead{$\lambda_\odot$} &
\colhead{$\lambda_{\odot,\mathrm{act}}$} &
\colhead{$\alpha_g$} &
\colhead{$\delta_g$} &
\colhead{$v_g$} &
\colhead{$\Delta\psi$} &
\colhead{$a$} &
\colhead{$q$} &
\colhead{$e$} &
\colhead{$\omega$} &
\colhead{$\Omega$} &
\colhead{$i$} &
\colhead{$D_H$}
}
\startdata
July Fornax--Eridanus (JFE) & 110.0 & 108--112 & 54.4 & $-30.6$ & 48.7 & \nodata & 3.42 & 0.898 & 0.710 & 316.9 & 290.0 & 91.3 & \nodata \\
M2024-N1 & 102.7 & 99.7--104.0 & 45.3 & $-38.3$ & 52.0 & 10.7 & 31.4 & 0.990 & 0.962 & 341.5 & 282.7 & 92.9 & 0.468 \\
Fornacids (FOR) & 116.1 & \nodata & 37.4 & $-31.1$ & 57.3 & 14.6 & 17.5 & 1.01 & 0.942 & 8.8 & 296.1 & 108.0 & 0.817 \\
$\psi$-Phoenicids (PPH) & 111.2 & 98--120 & 30.0 & $-47.1$ & 36.3 & 24.9 & 1.26 & 0.889 & 0.293 & 63.5 & 291.0 & 74.6 & 0.945 \\
July Phoenicids (PHE) & 110.5 & 101--119 & 30.3 & $-48.0$ & 42.6 & 25.3 & 1.34 & 0.934 & 0.670 & 41.4 & 290.5 & 80.9 & 0.943 \\
2021 BB2 & \nodata & \nodata & \nodata & \nodata & \nodata & \nodata & 1.6958 & 1.0699 & 0.36910 & 278.9896 & 303.0680 & 63.1603 & 0.730 \\
C/2004 B1 (LINEAR) & \nodata & \nodata & \nodata & \nodata & \nodata & \nodata & $-1733.1$ & 1.58751 & 1.000916 & 328.0187 & 273.3453 & 113.8153 & 0.661 \\
\hline
\enddata
\end{deluxetable*}

\section{Discussion}
\label{sec:discussion}

The preceding sections show that PANSY is best interpreted as a stable, nearly continuously sampled southern head-echo survey. The unchanged radar mode and Antarctic viewing geometry make the catalogue especially useful for comparing radiant morphology, seasonal activity, and orbit structure across southern sporadic and shower populations. At the same time, the sparse-array illumination pattern and speed-dependent radar selection mean that raw detections cannot be read directly as intrinsic meteoroid fluxes, without further debiasing. The discussion therefore focuses on what can be inferred robustly from the present catalogue: source morphology, height-dependent population differences, relative source rates, shower associations, and the prospects for improving mass constraints.

The approximately sinusoidal annual variation in the detected rate (Figure~\ref{fig:counts_solar_longitude}) is primarily a consequence of seasonal changes in radiant visibility at Syowa rather than an equivalent modulation of the incident meteoroid flux. The broad minimum from August through December occurs when the apex sources reach large zenith angles and are partly blocked by the horizon, whereas the maximum from February through May occurs when the apex complex is highest above the horizon. The apex sources dominate meteor head echo survey raw counts because their meteoroids are predominantly on retrograde orbits and consequently encounter the atmosphere at high speed. To first order, the aerodynamic energy-loss rate available for heating and ablation scales as
\begin{equation}
 \left|\frac{dE}{dt}\right| \propto \rho_a r^2 v^3 ,
\end{equation}
where $\rho_a$ is the atmospheric mass density, $r$ is the meteoroid radius, and $v$ is its speed. The associated increase in ablation rate and head-echo plasma density allows the radar to reach lower limiting masses at higher speeds \citep{close_2007_meteor,dimant2017formation}. This speed-selection effect is also apparent in the velocity histogram above Figure~\ref{fig:mass_speed_bounds}: fast meteoroids dominate the raw detections, while the slower population becomes dominant after the approximate velocity-sensitivity correction is applied.

Compared with optical meteor surveys, the head-echo radiant distribution is considerably more diffuse, as previously reported for the MU radar by \citet{sato2000mu}. This difference is attributed to the smaller meteoroids sampled by head-echo radars: Poynting--Robertson drag causes their orbits to evolve differently and disperse more rapidly than those of the larger meteoroids detected by optical surveys \citep{wiegert2009dynamical}. Nevertheless, many shower enhancements are sufficiently well defined to be recognized by direct visual inspection of the radiant maps. These include SDA, PAU, CAP, DSX, STA, ORI, OHY, AAN, ETA, OCE, SMA, SSG, DLT, NZC, and SZC, as well as the Working List showers OCD, RPH, ZPH, PPH, ACD, and AAN. Thus, although the diffuse sporadic background dominates much of the head-echo radiant distribution, coherent shower structure remains readily visible without a dedicated stream-search algorithm.

\subsection{Narrow apex source}

One of the distinct features in the radiant distribution is the narrow apex source. A zonally narrow and meridionally wide component within the northern and southern apex sources was first reported in Jicamarca head-echo observations by \citet{chau2007jicamarca_sources}, but it has not been detected in specular meteor-radar surveys. In the PANSY catalogue, the narrow apex is visible in both the raw counts and the debiased radiant distributions as a compact enhancement near the Earth's direction of motion (Figure~\ref{fig:radiant_distribution}).

To isolate the narrow apex source, we applied a spherical-harmonic low-pass filter, primarily in the zonal direction, to the radiant distribution. The resulting low-spatial-frequency component is shown in panel c) of Figure~\ref{fig:height_band_radiants}; panel d) shows the high-spatial-frequency residual obtained by subtracting the smooth background radiant distribution in panel c) from the raw counts. This decomposition further suggests that the narrow apex source is a coherent, meridionally extended structure that peaks near $\beta=-15^\circ$. The structure is not confined to the apex region, but extends into the southern toroidal source.

The decomposed narrow-apex source closely resembles the central apex condensations modeled by \citet{wiegert2009dynamical}. Their simulations associate this morphology with old ($\gtrsim10^5$~yr) retrograde meteoroids having radii $r\lesssim100\,\micron$, originating from 55P/Tempel--Tuttle or a dynamically similar Halley-type parent with perihelion distance near 1~AU. Radiation forces and planetary perturbations evolve this material onto Earth-crossing orbits.

\subsection{Initial height distribution}

\begin{figure*}
\centering
\includegraphics[width=\textwidth]{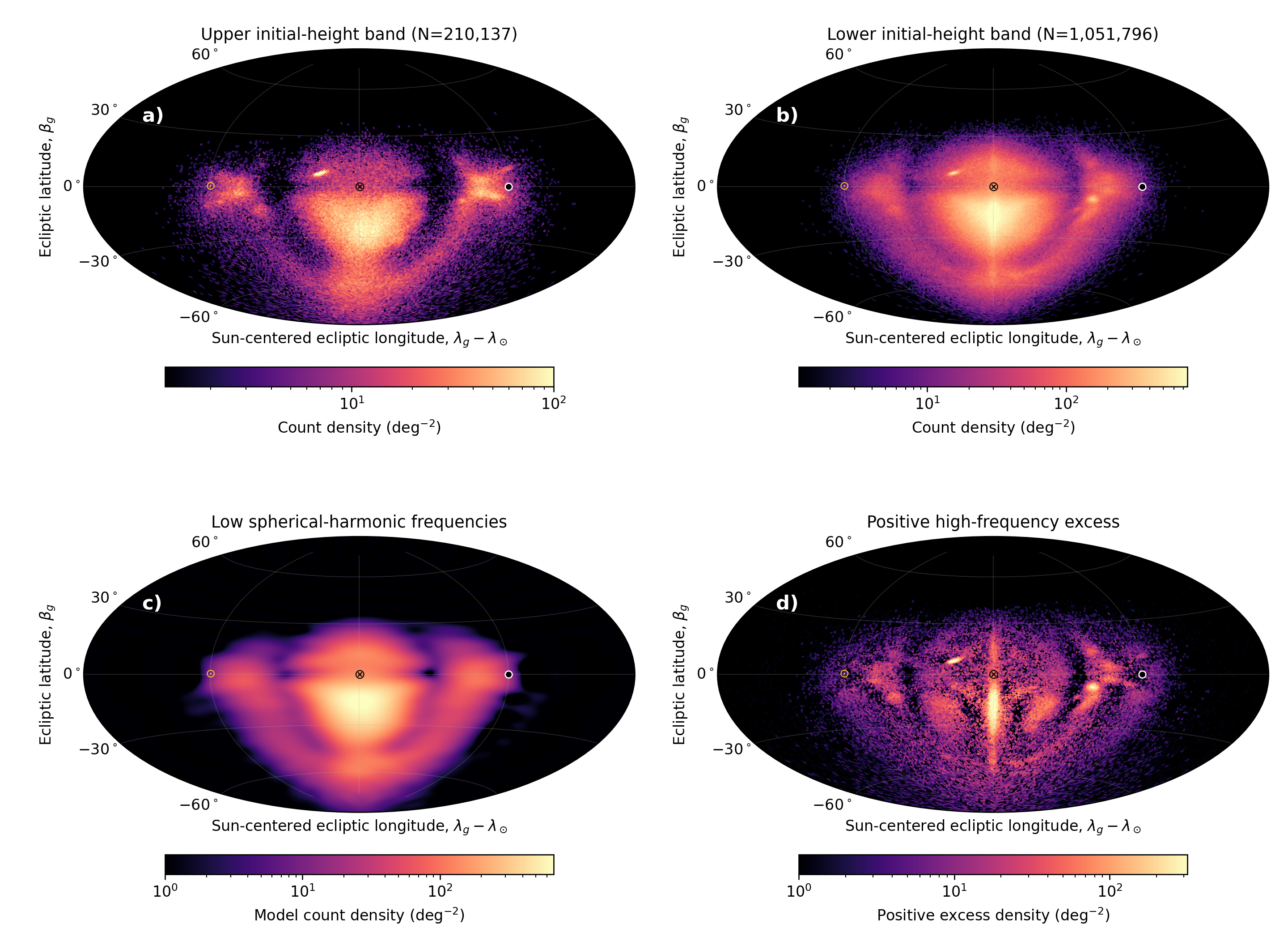}
\caption{Sun-centered ecliptic radiant density decomposed in four ways. Panels a) and b) show the radiant distributions of meteors in the upper and lower initial-height bands in Figure~\ref{fig:height_velocity}. Panel c) shows the low-spatial-frequency component of the raw radiant counts in Figure~\ref{fig:sporadic_source_apertures}, obtained by spherical-harmonic smoothing primarily in the zonal direction. Panel d) shows the high-spatial-frequency component obtained by subtracting panel c) from the raw counts. This component isolates the narrow apex source.}
\figureprovenance{height\_band\_web\_gui.py; plot\_height\_band\_spatial\_frequency\_decomposition.py}
\label{fig:height_band_radiants}
\end{figure*}

The initial-height distribution in Figure~\ref{fig:height_velocity} shows a clear two-banded structure. To investigate whether the meteors in these bands have different origins, Figure~\ref{fig:height_band_radiants} compares their Sun-centered radiant densities in panels a) and b). The upper-height band has more structured helion and antihelion source regions, a narrower ring around the apex, and a wider and deeper minimum between the apex direction and that ring. The lower-height band has a broader apex complex, and the narrow apex feature is visible only in this lower-band radiant distribution. The $\eta$ Aquariid shower region is also more prominent in the upper band than in the lower band.

These differences are consistent with the idea that the two height bands are not only an instrumental threshold effect, but also reflect a mixture of compositional and size-dependent meteoroid properties. Differential ablation provides a plausible physical mechanism and has been observed in high-power radar measurements of micrometeoroids \citep{janches2009differential}. Chemical-ablation modelling predicts that volatile alkalis such as Na and K are released above the main Fe, Mg, and Si injection region, whereas refractory Ca peaks lower \citep{vondrak2008chemical}. Pyrolysis of organic material can also occur at comparatively low particle temperatures and  contribute significantly to early high-altitude mass loss  \citep{bones2022organic}. The higher-altitude band could therefore be associated with volatile release and organic pyrolysis, while the lower band could trace the main ablation of less volatile meteoric metals.

The effects of size and differential ablation cannot be disentangled here. Larger meteoroids are also detected at higher altitude because they cross the radar threshold earlier, but without a double-peaked differential-ablation height profile, that size-threshold effect would more naturally broaden or continuously shift the onset-height distribution; by itself, it does not obviously explain two separated bands. The interpretation remains qualitative because onset height is also shaped by radar sensitivity, range, entry geometry, meteoroid mass, velocity, and composition \citep{close_2007_meteor}; related altitude-dependent head-echo effects have been discussed in the EISCAT--MAARSY comparison by \citet{huyghebaert2025altitude}.

The absence of the narrow apex feature in the upper initial-height-band radiant distribution (Figure \ref{fig:height_band_radiants}a) may also be physically meaningful. As discussed earlier, \citet{wiegert2009dynamical} interpret the narrow apex source as $>10^5$-year-old retrograde material related to 55P/Tempel--Tuttle or a similar Halley-type comet. Such material would have been space-weathered for much longer than typical meteor sources, contributing to the loss of organic materials and volatile compounds such as Na and K in space, long before atmospheric entry. In that case, its preferential appearance in the lower band would be qualitatively consistent with a more volatile-depleted composition. 

\subsection{Relative sporadic-source rates}

The morphology of the radiant distribution is consistent with earlier head-echo radar surveys \citep{chau+woodman-2004,kero_mu_catalogue,schult_2018_meteoroid}. After the exposure, zenith-angle, and speed corrections are applied, the largest relative rates lie close to the ecliptic in the helion and antihelion directions, as seen also with the CMOR specular meteor radar survey \citep{campbell_brown_2008_sporadic_radiants}. A local minimum extends along the ecliptic across all longitudes, possibly because this region of orbital-element space has a higher probability of planetary encounters, which would remove meteoroids occupying this part of the orbital elements space.

As a compact summary of the sporadic-source morphology, we integrated the radiant maps over meteor source apertures. These apertures are intended as diagnostic regions, not as a formal sporadic-source decomposition. The helion and antihelion apertures use $|\lambda'_g| \leq 45^\circ$ and $|\lambda'_g-180^\circ|\leq45^\circ$, respectively, both with $|\beta|\leq39.5^\circ$. The apex aperture uses $|\lambda'_g-270^\circ|\leq45^\circ$ and $|\beta|\leq39.5^\circ$. The southern toroidal aperture uses $|\lambda'_g-270^\circ|\leq69.5^\circ$ and $-80^\circ\leq\beta\leq-42.5^\circ$.

\begin{figure*}
\centering
\includegraphics[width=\linewidth]{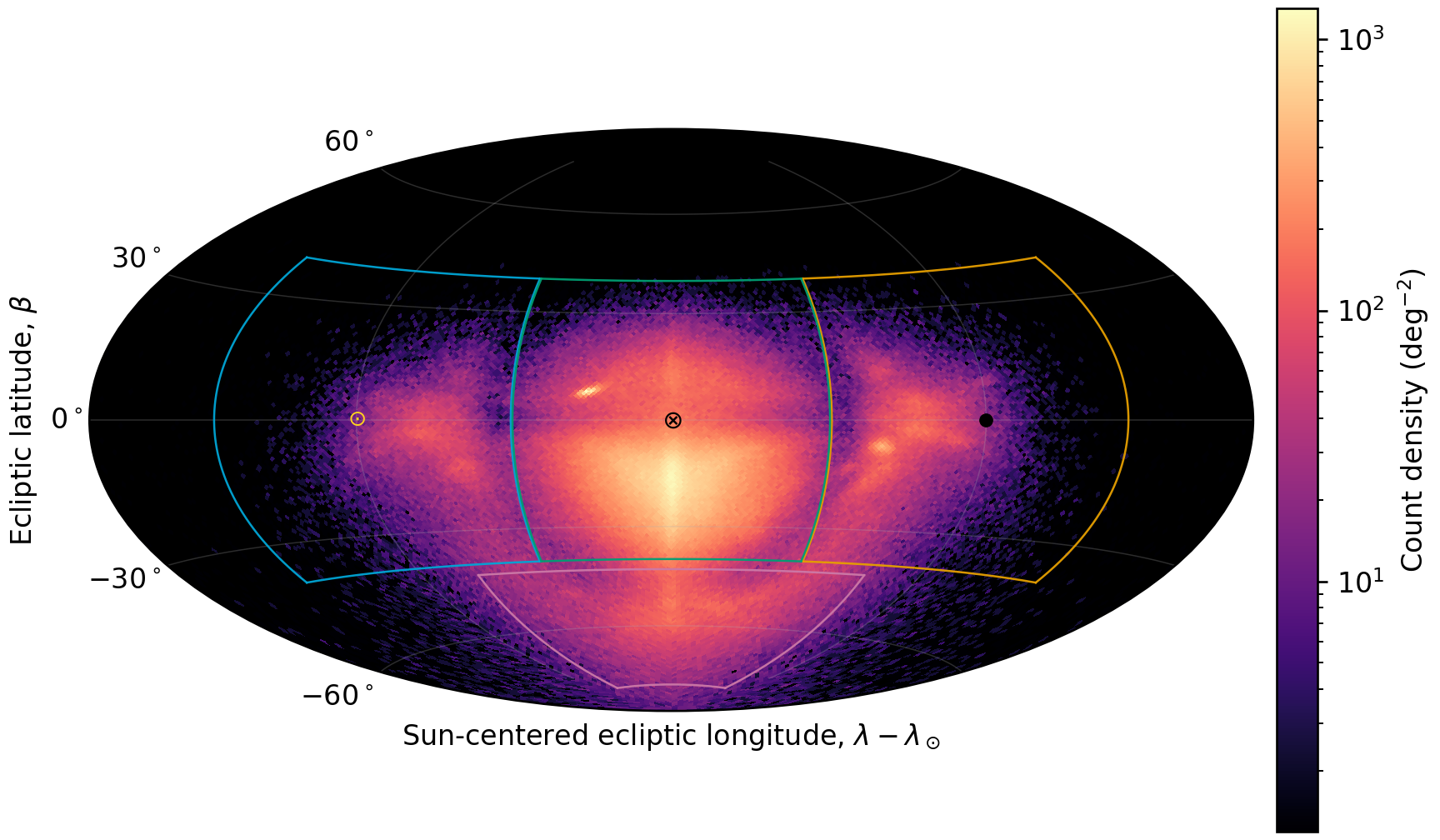}
\caption{Diagnostic sporadic-source apertures used for the relative-rate calculation in Table~\ref{tab:sporadic_source_rates}. The cyan, orange, green, and magenta bounding boxes identify the helion, antihelion, apex, and southern toroidal regions, respectively.}
\figureprovenance{plot_sporadic_source_annotations.py}
\label{fig:sporadic_source_apertures}
\end{figure*}

Figure~\ref{fig:sporadic_source_apertures} shows these apertures on the raw-count map. For the apex aperture we further separate the broad apex contribution from the narrow apex enhancement using the smoothed and residual maps shown in Figure~\ref{fig:height_band_radiants}. The same smooth/residual split is applied to the helion, antihelion, and southern toroidal apertures only as a diagnostic indication of how much compact foreground structure is superposed on each broad source. Table~\ref{tab:sporadic_source_rates} gives three normalizations: raw counts, exposure-plus-zenith-angle-corrected aperture rates, and the same rates with an additional $1/v_g^3$ speed weight that approximately accounts for the first-order velocity dependence of plasma production.

The two corrected estimates account for the exposure time and zenith angle of each radiant, but their integration over the full catalogue implicitly treats the source fluxes as stationary through the year. Figure~\ref{fig:radiant_snapshots} shows that this approximation is poor: compact shower radiants appear and disappear with solar longitude, and even the broad toroidal source varies seasonally in both flux and radiant morphology \citep{campbell_brown_wiegert_2009_toroidal}. The fractions in Table~\ref{tab:sporadic_source_rates} are therefore averages over the particular temporal sampling of this catalogue, not universal source fractions. More broadly, the prevalence of time-dependent substructure suggests that the nominally sporadic complex is not completely phase mixed and that many meteoroids may retain dynamical memory of parent populations or evolved streams \citep{campbell_brown_wiegert_2009_toroidal,wiegert2009dynamical}. This does not imply that every individual sporadic meteor can be associated with a known parent, but it cautions against treating the sporadic background as a smooth, invariant population.

The raw and aperture-corrected fractions are dominated by apex-region detections: the smooth and structured apex components together account for 72.9\% of the raw counts and 68.1\% of the aperture-corrected rate. The structured narrow apex component comprises 14.9\% of raw apex detections and 21.4\% of the aperture-corrected apex rate. With the velocity weighting included, the helion and antihelion apertures instead dominate the relative flux, contributing 35.4\% and 42.6\%, respectively, or 78.0\% together. Their structured components account for 17.1\% and 20.2\% of the total corrected rate. The smooth and narrow apex components contribute 14.3\% and 5.6\%, while the smooth and structured southern toroidal components contribute 1.7\% and 0.4\%. These numbers should not be read as global meteoroid-flux fractions, because the apertures omit diffuse background and overlap only approximately with the physical source populations. They nevertheless show that correcting the leading observational biases changes the inferred picture dramatically: the helion and antihelion sources together account for 78\% of the debiased Earth-intercepting flux. This agrees with models of the meteoric complex, which attribute 85\% of the terrestrial dust mass influx to Jupiter-family-comet particles \citep{nesvorny2010cometary,nesvorny2011dynamical,soja2019imem2}.

The velocity-weighted antihelion-to-helion ratios are 1.22 for the smooth component, 1.18 for the structured component, and 1.20 for all counts. These values should not be interpreted as unbiased intrinsic ratios of the two sporadic sources. Because the helion and antihelion radiants lie at low ecliptic latitude, their seasonal visibility and elevation are sampled unevenly from the high-latitude PANSY site; residual exposure and direction-dependent sensitivity errors can therefore affect their relative integrated rates. Meteor showers and other structured foregrounds also contribute strongly within these broad apertures and may shift the ratio away from that of the smooth underlying components. The measured ratios agree in sense with the mass-dependent asymmetry in the dynamical sporadic-meteoroid model of \citet{wiegert2009dynamical}. Their model predicts ratios of order 1.2--1.3 for smaller radar meteoroids, while the asymmetry weakens for meteoroids with radii larger than about $250\,\mu$m. In that model the asymmetry is not primarily an observing-selection effect, but is produced by the size-dependent delivery and orbital evolution of young Jupiter-family-comet meteoroids, especially particles from 2P/Encke or an orbitally similar source, under radiation pressure and Poynting--Robertson drag over the last $10^3$--$10^4$~yr.

\begin{table}
\centering
\caption{Relative source fractions in diagnostic sporadic-source apertures. Percentages are normalized over the four source totals; the indented smooth and structured rows are subdivisions that sum to their source total. The final three rows give antihelion-to-helion ratios for the smooth, structured, and combined components.}
\label{tab:sporadic_source_rates}
\footnotesize
\begin{tabular}{@{}lccc@{}}
\hline
Region & Raw & Aperture & Aperture + $v_g$ \\
 & counts & debiased & debiased \\
\hline
\textbf{Helion, total} & \textbf{6.1\%} & \textbf{13.8\%} & \textbf{35.4\%} \\
\quad Smooth & 4.3\% & 8.5\% & 18.3\% \\
\quad Structured & 1.8\% & 5.3\% & 17.1\% \\
\textbf{Antihelion, total} & \textbf{10.1\%} & \textbf{15.9\%} & \textbf{42.6\%} \\
\quad Smooth & 7.0\% & 10.0\% & 22.5\% \\
\quad Structured & 3.1\% & 5.8\% & 20.2\% \\
\textbf{Apex, total} & \textbf{72.9\%} & \textbf{68.1\%} & \textbf{19.9\%} \\
\quad Smooth & 62.0\% & 53.6\% & 14.3\% \\
\quad Narrow, structured & 10.8\% & 14.6\% & 5.6\% \\
\textbf{S. toroidal, total} & \textbf{11.0\%} & \textbf{2.2\%} & \textbf{2.1\%} \\
\quad Smooth & 9.1\% & 1.8\% & 1.7\% \\
\quad Structured & 1.8\% & 0.4\% & 0.4\% \\
\hline
Antihelion/helion, smooth & 1.63 & 1.18 & 1.22 \\
Antihelion/helion, structured & 1.74 & 1.10 & 1.18 \\
Antihelion/helion, all & 1.66 & 1.15 & 1.20 \\
\hline
\end{tabular}
\end{table}

\subsection{The CAP--DCS complex}
\label{sec:shower_discussion}

The earlier literature treated DCS as a single CAP-related daytime shower. \citet{brown_etal_2010_cmor_minor_showers} reported a compact radar enhancement at $\lambda_\odot=301.0^\circ$, with $(\alpha_g,\delta_g)=(304.7^\circ,-29.2^\circ)$ and $v_g=23.8\,\kms$, and interpreted it as the likely daytime counterpart of CAP. \citet{jenniskens_vaubaillon_2010_capricornids} independently modeled meteoroids released from 169P/NEAT and predicted an opposite-node shower at $(\alpha_g,\delta_g)=(309.0^\circ,-29.4^\circ)$ and $v_g=22.7\,\kms$.

The PANSY daytime radiant extends across the activity interval rather than separating cleanly into independent streams. Its activity profile in Figure~\ref{fig:capricornid_stream} contains a localized enhancement near $\lambda_\odot=313^\circ$, but the uneven exposure around this longitude prevents the neighboring minimum from being measured robustly. The reference orbit of 169P/NEAT has $a=2.60$~AU, corresponding to an orbital period of approximately 4.2~yr. If the localized enhancement is real, it could represent meteoroids released during a different perihelion passage or fragmentation epoch from the material forming the broader DCS radiant. We therefore interpret it as possible localized flux structure within one extended DCS radiant rather than evidence for a separate shower. This interpretation is also consistent with the nearly equal observed durations of DCS and CAP: approximately $34^\circ$ and $33^\circ$ in solar longitude, respectively, at opposite nodes.

The measured CAP and DCS orbits occupy the same broad region of orbital-element space as 169P/NEAT, but that similarity alone does not demonstrate a common origin. Table~\ref{tab:dcs_mean_parameters} gives $D_H=0.180$ for the PANSY DCS mean and $D_H=0.172$ for the PANSY CAP mean relative to the present 169P/NEAT orbit. The largest angular differences occur in $\omega$ and $\Omega$, the elements changed by apsidal and nodal precession. Substantial evolution is expected for millennia-old 169P/NEAT ejecta under planetary perturbations, particularly Jovian perturbations \citep{jenniskens_vaubaillon_2010_capricornids}.

The dynamical model of \citet{jenniskens_vaubaillon_2010_capricornids} predicts that CAP activity will increase strongly as the nodes of the modeled 169P/NEAT meteoroid population evolve toward Earth's orbit, with the largest enhancement in the 23rd and 24th centuries. The modeled peak rate depends on assumptions about shower duration and particle evolution, so its magnitude remains uncertain. Extending the model to the smaller particles sampled by PANSY will require inclusion of size-dependent radiation pressure and direct comparison with the measured CAP and DCS activity profiles.

\subsection{July Fornax--Eridanus shower candidate}

The compact radiant, recurrence in both observing years, and fivefold excess over the adjacent-interval background support interpreting JFE as a shower candidate rather than a chance concentration of sporadic meteors. Its status remains provisional until the stream is recovered independently or identified in another orbit catalogue.

The nearest Working List radiant is M2024-N1, but its mean peak occurs $7.3^\circ$ earlier in solar longitude and its radiant is separated from JFE by $10.7^\circ$; its mean speed is also $3.3\,\kms$ higher and its orbital dissimilarity is $D_H=0.468$. FOR is the next-nearest radiant at $14.6^\circ$, while PPH and PHE are approximately $25^\circ$ away. The complete Working List comparison therefore provides no convincing identification of JFE with a listed shower.

We also searched the Minor Planet Center near-Earth-asteroid and comet catalogues downloaded on 2026 July 27 using the Jopek $D_H$ orbital-dissimilarity criterion \citep{jopek1993remarks}. Even the nearest asteroid and comet have large $D_H$ values, so the present search identifies no convincing parent body.


\subsection{Pulse-to-pulse Doppler tests for mass constraints}
\label{sect:p2p}

For the production catalogue, radial velocity is currently estimated independently within each transmitted pulse from the peak of an FFT-based range--Doppler matched-filter bank. These within-pulse estimates constrain the meteoroid entry speed, but their pulse-to-pulse scatter is typically too large to separate changes in radial velocity caused by geometric motion across the radar field of view from deceleration caused by atmospheric drag. For this reason, the current catalogue release does not include initial radius or mass estimates; it includes peak signal-to-noise ratio as the primary amplitude-related quantity.

Inter-pulse phase change can be used to determine radial velocity more accurately, to within a few tens of metres per second when the integer phase ambiguity is resolved \citep{chau+woodman-2004}. This has been demonstrated with the MU radar at a 3.12~ms sampling interval \citep{kero_2012_meteor}. PANSY revisits a given transmit-beam direction every 8~ms, substantially longer than the interpulse period used in earlier studies \citep{schult2013determination,kero_2012_meteor}. At the PANSY wavelength this gives adjacent phase-velocity aliases separated by only about $400~\mathrm{m\,s^{-1}}$, so a robust estimator must resolve the correct alias as well as measure the phase accurately.

\begin{figure}
\centering
\includegraphics[width=\linewidth,height=0.62\textheight,keepaspectratio]{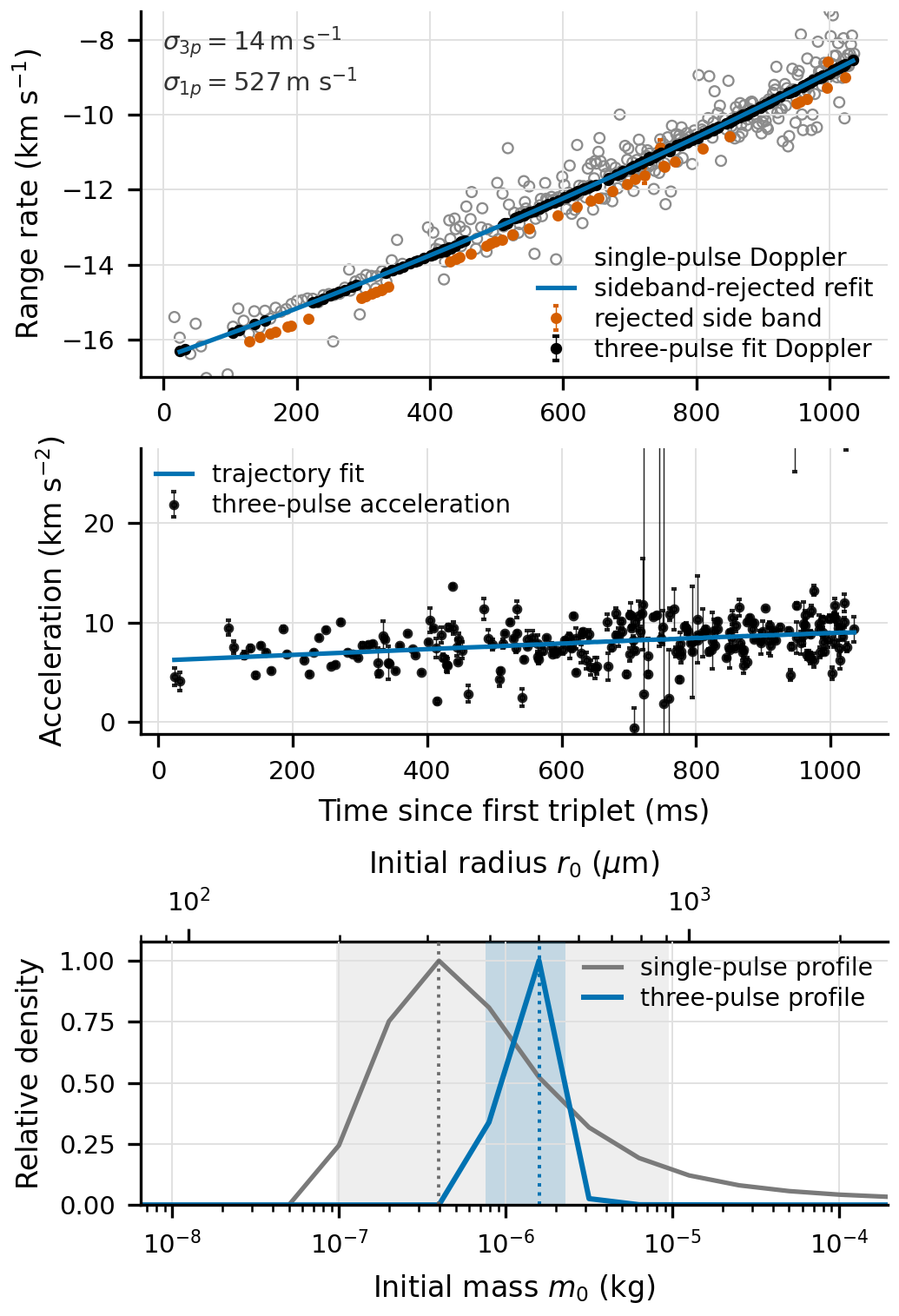}
\caption{Pulse-to-pulse Doppler diagnostic for the same meteor event shown in Figure~\ref{fig:example_detection}. The upper panel compares single-pulse Doppler estimates with the three-pulse complex-voltage Doppler fit. Red points are rejected sideband triplets, identified from the initial trajectory-fit residual using $|\Delta v|\geq200~\mathrm{m\,s^{-1}}$; the blue curve is a refit to the remaining main velocity branch. The quoted standard deviations compare the single-pulse Doppler residuals with the post-refit three-pulse residuals. The middle panel shows the corresponding three-pulse radial acceleration estimates. The lower panel compares the marginal $r_0$ and $m_0$ profiles before and after adding the cleaned three-pulse Doppler and acceleration measurements, showing the resulting shrinkage of the mass interval.}
\figureprovenance{plot\_pulse\_to\_pulse\_doppler\_example.py}
\label{fig:pulse_to_pulse_doppler_example}
\end{figure}

\begin{figure}
\centering
\includegraphics[width=\linewidth]{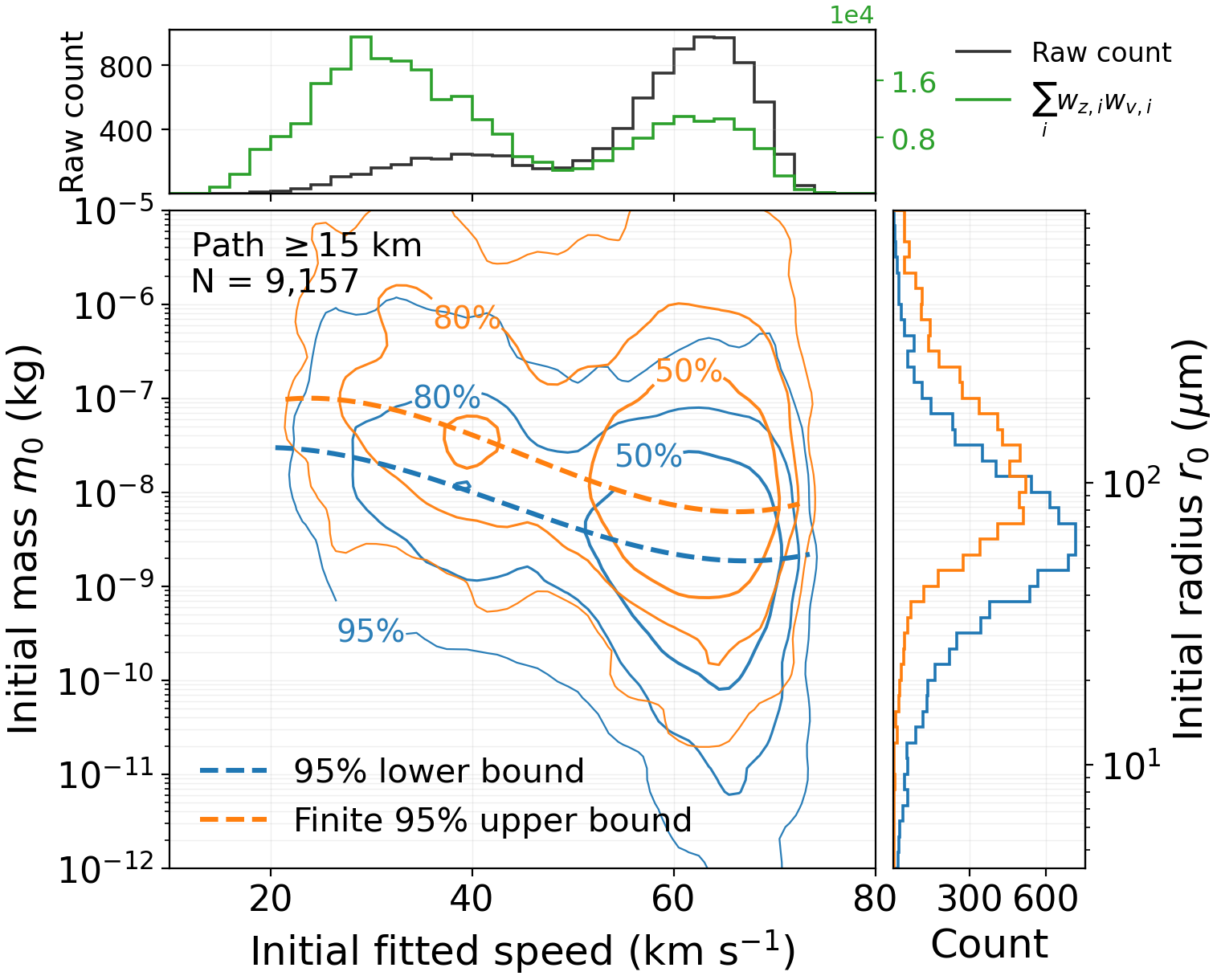}
\caption{Initial dynamic-mass bounds for a representative subset of 9,157 meteors with observed path lengths greater than 15~km. The main panel shows bound density as a function of fitted initial speed, the attached right panel shows the corresponding marginal mass histograms, and the top panel shows the raw velocity count (black) together with the empirically debiased count $\sum_i w_{z,i}w_{v,i}$ (green). The latter uses $w_{z,i}=[\max(\cos z_{R,i},0.15)]^{-1.96}$ and $w_{v,i}=(73~\mathrm{km\,s^{-1}}/v_{0,i})^3$. Blue denotes 95\% lower bounds and orange denotes finite 95\% upper bounds. The dashed curves trace smoothed conditional modes of the two bound distributions as a function of speed. The outer ordinate gives the corresponding initial radius for the adopted meteoroid density.}
\figureprovenance{summarize\_catalogue\_mass\_profiles.py}
\label{fig:mass_speed_bounds}
\end{figure}

In order to investigate the feasibility of obtaining improved Doppler velocity and initial radius estimate, we implemented an experimental estimator that fits the complex echo phase jointly over three successive observations of the same beam direction. We first demonstrate the method with the individual meteor event in Figure~\ref{fig:pulse_to_pulse_doppler_example}. The pulse-to-pulse model uses the raw voltage of three consecutive echoes. For each echo triplet, the calibrated module voltages are coherently combined using the selected angle of arrival and aligned to the fractional-sample matched-filter range. The complex pulse sequence is fitted locally as
    \begin{equation}
     \begin{split}
     z_{k,b}={}&A_k E_{k,b}\exp\!\left[i\Phi(t_{k,b})\right]+\epsilon_{k,b},\\
     \Phi(t)={}&\phi_0+\frac{4\pi}{\lambda}v_{R,0}(t-t_0)\\
     &+\frac{2\pi}{\lambda}a_R(t-t_0)^2 .
     \end{split}
     \label{eq:pulse_to_pulse_doppler}
\end{equation}
Here $z_{k,b}$ is the decoded, SNR-weighted complex voltage for baud $b$ of pulse $k$, and $E_{k,b}$ is its known decoded envelope. The separate real amplitudes $A_k$ allow the echo strength to vary between pulses. The reference time $t_0$ is the leading edge of the first transmitted pulse, and the same continuous time coordinate is retained across the inter-pulse gaps. The initial radial velocity $v_{R,0}=\dot R(t_0)$, radial range acceleration $a_R=\ddot R$, and phase $\phi_0$ are shared by the triplet. The implementation used for Figures~\ref{fig:pulse_to_pulse_doppler_example} and \ref{fig:mass_speed_bounds} includes both 8~ms and 16~ms pulse-to-pulse phase acceleration observables.

Because the inter-pulse phase is measured modulo $2\pi$, equally spaced pulses admit discrete velocity and acceleration branches. For a same-beam pulse spacing $T$, their separations are
\begin{equation}
 \Delta v_{\rm amb}=\frac{\lambda}{2T},
 \qquad
 \Delta a_{\rm amb}=\frac{\lambda}{2T^2}.
 \label{eq:pulse_to_pulse_ambiguities}
\end{equation}
For $\lambda=6.379$~m and $T=8$~ms, these are $398.7~\mathrm{m\,s^{-1}}$ and $49.8~\mathrm{km\,s^{-2}}$, respectively. The initial FFT Doppler estimate identifies the relevant velocity and acceleration neighborhood, after which the candidate phase branches are compared in the three-pulse fit.

Figure~\ref{fig:pulse_to_pulse_doppler_example} shows pulse-to-pulse Doppler diagnostics for the same event as Figure~\ref{fig:example_detection}. The single-pulse Doppler estimates have the expected larger scatter, while the three-pulse complex-voltage Doppler fit follows the trajectory range-rate solution and the three-pulse phase evolution provides a direct local acceleration estimate. For this event, selecting the main three-pulse velocity branch by requiring the residual from the initial trajectory fit to be less than $200~\mathrm{m\,s^{-1}}$, and then refitting that branch, gives a residual standard deviation of $14~\mathrm{m\,s^{-1}}$. The corresponding single-pulse Doppler residual standard deviation is $527~\mathrm{m\,s^{-1}}$.

The measured positions and selected range-rate observables are compared with the numerical solution of Equations~\ref{eq:mass_radius}--\ref{eq:ceplecha_ablation}. A marginal probability density estimate for the initial radius $r_0$ is computed by holding only the initial value $r_0$ fixed on a logarithmic grid and refitting the remaining state parameters. Radius continues to evolve according to the ablation equation in every profile fit. Equation~\ref{eq:mass_radius} maps this initial-radius distribution to an initial dynamic-mass distribution after adopting $\rho_m$.

Occasional velocity side bands are nevertheless visible in this processing. These side bands are separated by about $400~\mathrm{m\,s^{-1}}$, corresponding to the pulse-to-pulse $2\pi$ phase ambiguity. The ambiguity is usually resolved by the trajectory and local likelihood constraints, but not always. Pulse-to-pulse phase progression therefore has the potential to substantially improve dynamic trajectory fitting, but robust side-band rejection must be solved before it can be used as a high-quality Doppler estimator throughout the catalogue.

The example also illustrates why the pulse-to-pulse information is potentially important for dynamic mass estimation. With the original single-pulse Doppler information, the marginal $r_0$ profile retains a long high-mass tail, because the measured arc only weakly distinguishes modest deceleration from nearly constant speed. Repeating the profile calculation with the cleaned three-pulse velocity branch and with acceleration outliers removed gives a bounded 95\% interval, $r_0=390$--$570\,\microm$ and $m_0=7.5\times10^{-7}$--$2.3\times10^{-6}$~kg for the adopted density. 

We applied the same three-pulse processing to 50,000 randomly chosen PANSY head echoes. Of these, 9,157 had observed path lengths greater than 15~km and yielded a 95\% initial-radius or mass constraint. Figure~\ref{fig:mass_speed_bounds} shows this representative-sample summary. It demonstrates that pulse-to-pulse voltage information can sharpen the deceleration observable beyond the standard within-pulse FFT range--Doppler products.

The high Doppler resolution subset provides an estimate of the typical initial radii and masses sampled by PANSY. In this subset, slow meteors with $v_0=30$--$45~\kms$ have median 95\% lower and finite-upper bounds of approximately $r_0=100\,\microm$ and $r_0=170\,\microm$, corresponding to $m_0=1.2\cdot10^{-8}$ and $6.5\cdot10^{-8}$~kg for $\rho_m=3000~\mathrm{kg\,m^{-3}}$. Fast meteors with $v_0=60$--$70~\kms$ are constrained to smaller radii, with corresponding median bounds of about $r_0=50\,\microm$ and $r_0=90\,\microm$, or $m_0=1.6\cdot10^{-9}$ and $1.0\cdot10^{-8}$~kg. These values imply that the survey is most sensitive to meteoroids with radii of order $100\,\microm$. However, the uncertainty is still rather large, nearly one order of magnitude in mass.

The improved processing however is not yet a mature catalogue-scale mass-estimation pipeline. Extending the improved Doppler fitting to the full catalogue requires a more robust and numerically efficient analysis, and therefore initial radius and mass estimates are not yet included for all meteoroids within the catalogue.


\section{Conclusions}
\label{sec:conclusions}

The PANSY observations provide a nearly continuous Antarctic head-echo orbit catalogue spanning all solar longitudes. The catalogue contains two million fitted meteors, with extensive coverage of southern ecliptic radiants from the PANSY location at $69^\circ$ S. Analysis of a long-duration subset indicates sensitivity to meteoroids with radii of order $100\,\mu\mathrm{m}$.

The initial detection-height distribution is double-banded. Such structure is predicted by chemical ablation models, in which more volatile constituents ablate at lower temperatures and therefore at higher altitudes \citep{vondrak2008chemical}. The upper and lower bands have distinct radiant distributions and thus sample different orbital populations, suggesting differences in their physical properties that may include chemical composition. The narrow apex source, for example, is observed only in the lower band, which could reflect depletion of volatile constituents through long-term space weathering. Meteoroid ablation modeling will be needed to determine whether differential ablation reproduces the observed bands and whether catalogue-wide height statistics can constrain models of meteoroid chemical composition.

The continuous PANSY measurements resolve the annual variability of both sporadic radiant sources and meteor showers. Because the same experiment and analysis are used throughout the catalogue, the relative fluxes of different sporadic sources can be compared consistently. The measured antihelion-to-helion flux ratio of approximately 1.2 agrees with previous work discussed by \mbox{\citet{wiegert2009dynamical}}. The DCS--CAP case demonstrates that compact showers near the ecliptic can be resolved, while the July Fornax--Eridanus shower candidate demonstrates the potential of the survey to discover previously unrecognized meteor showers.

No systematic search for minor meteor showers has yet been conducted with the PANSY catalogue. Nevertheless, based on visual inspection of the radiant distributions, the catalogue contains several compact enhancements that appear to be minor showers not currently included in the IAU Meteor Data Center Working List. Such a search is planned to be conducted once the full observing record has been compiled following the planned conclusion of the survey in September 2027, when recurrent activity can be tested over a longer time interval.

The PANSY catalogue is complementary to continuous MAARSY head-echo observations at a similar northern polar latitude \citep{schult_2017_results,schult_2018_meteoroid}. A future joint analysis of the Antarctic and Arctic catalogues would provide nearly global radiant-latitude coverage with closely related HPLA observing techniques, allowing hemispheric selection effects and recurrent shower detections to be tested directly.

For typical detected radii of approximately $100\,\microm$, atmospheric deceleration over a head-echo arc is small. The single-body deceleration fit can often reject smaller bodies but cannot determine an upper bound on radius. This initial catalogue release will therefore not include estimates of initial radius. Future signal-processing improvements will aim to reduce the Doppler uncertainty and thereby improve constraints on initial radius, as discussed in Section~\ref{sect:p2p}. In particular, the pulse-to-pulse phase method demonstrated here will be developed into a robust Doppler-shift estimator with automatic phase-alias resolution and applied to the full catalogue. The temporarily retained level~1 complex-voltage cuts preserve the information required for this reprocessing while they remain available.

Another priority is an empirical characterization of the transmit beam pattern and its sidelobes, enabling accurate radar-cross-section estimates that can provide an independent mass estimate \citep{tarnecki_2021_meteoroid}. As Figure~\ref{fig:meteor_positions} demonstrates, the spatial occurrence distribution of accurately reconstructed meteor trajectories traces the main lobe and sidelobes of the transmitted beam. Using the position and SNR observables of meteors within the catalogue can possibly permit the sidelobe structure to be estimated directly from the meteor observations, including departures from the ideal analytic array model. This approach has already been demonstrated with MAARSY, where meteor head-echo occurrence maps were used to characterize beam pointing and width and to diagnose changes in the active array \citep{renkwitz_2017_antenna}. An empirical PANSY beam model would improve gain and radar-cross-section estimation along meteor trajectories.

PANSY radar operations are expected to continue until September 2027, when operation of the high-power, large-aperture radar is scheduled to cease. Meteor head-echo observations are planned to continue throughout the remainder of the PANSY radar's operational lifetime, extending the catalogue's temporal coverage and improving the statistics of recurrent showers and time-variable radiant sources. A linear extrapolation of the 2025 January 26--2026 July 26 accumulation rate, assuming unchanged observing duty cycle, detection rate, and quality-selection fraction, predicts approximately 3.7 million fitted head echoes by the end of operations.

\begin{acknowledgments}
PANSY is a multi-institutional project with core members from the University of Tokyo, the National Institute of Polar Research, and Kyoto University. The PANSY radar was operated by the Japanese Antarctic Research Expedition (JARE). We thank the summer and wintering members of JARE and the PANSY operations and logistics teams at Syowa Station for their support. We thank Mr. K. Mushiake for assistance with the receiver testing and installation. This work was supported by JSPS KAKENHI Grant JP22H00169 (K.S.).

OpenAI Codex, using large language models, was used for writing and programming assistance during preparation of this manuscript. All generated material was reviewed and verified by the authors, who retain responsibility for the scientific content \citep{openai2025codex,vishniac2023chatbots}.
\end{acknowledgments}

\facility{PANSY}

\software{REBOUND, NRLMSISE-00}

\section*{Data Availability}

Version~1 of the PANSY Meteor Head-echo Orbit Catalogue is available through Zenodo \citep{juha_2026_21703650} as three archives. The level~2 and level~3 archives contain the individual meteor positions and orbits. Raw complex-voltage cuts are retained temporarily during analysis but are not intended for permanent preservation. The full level~1 collection is approximately 10~TB and therefore could not be included in the public Zenodo release. The published level~1 archive instead contains the complex-voltage cut for the example in Figure~\ref{fig:example_detection}. The catalogue can also be obtained through RadiantViz at \url{https://juha.no/radiantviz/}.

\bibliographystyle{aasjournalv7}
\bibliography{agusample}

@article{ceplecha1998meteor,
  author = {Ceplecha, Zden{\v e}k and Borovi{\v c}ka, Ji{\v r}{\'i} and Elford, W. Graham and ReVelle, Douglas O. and Hawkes, Robert L. and Porub{\v c}an, Vladim{\'i}r and {\v S}imek, Milo{\v s}},
  title = {Meteor Phenomena and Bodies},
  journal = {Space Science Reviews},
  year = {1998},
  volume = {84},
  number = {3--4},
  pages = {327--471},
  doi = {10.1023/A:1005069928850}
}

@article{kero2008determination,
  title={Determination of meteoroid physical properties from tristatic radar observations},
  author={Kero, Johan and Szasz, Csilla and Pellinen-Wannberg, Asta and Wannberg, Gudmund and Westman, Assar and Meisel, DD},
  journal={Annales Geophysicae},
  volume={26},
  number={8},
  pages={2217--2228},
  year={2008}
}

@article{de2008model,
  author = {{de Oliveira-Costa}, Ang{\'e}lica and Tegmark, Max and Gaensler, B. M. and Jonas, Justin and Landecker, T. L. and Reich, Patricia},
  title = {A Model of Diffuse Galactic Radio Emission from 10 MHz to 100 GHz},
  journal = {Monthly Notices of the Royal Astronomical Society},
  year = {2008},
  volume = {388},
  number = {1},
  pages = {247--260},
  doi = {10.1111/j.1365-2966.2008.13376.x}
}

@article{dimant2017formation,
  author = {Dimant, Y. S. and Oppenheim, M. M.},
  title = {Formation of Plasma Around a Small Meteoroid: 2. Implications for Radar Head Echo},
  journal = {Journal of Geophysical Research: Space Physics},
  year = {2017},
  volume = {122},
  number = {4},
  pages = {4697--4711},
  doi = {10.1002/2017JA023963}
}

@article{rein_liu_2012_rebound,
  author = {Rein, Hanno and Liu, Shang-Fei},
  title = {{REBOUND}: An Open-source Multi-purpose {N}-body Code for Collisional Dynamics},
  journal = {Astronomy \& Astrophysics},
  year = {2012},
  volume = {537},
  pages = {A128},
  doi = {10.1051/0004-6361/201118085}
}

@article{rein_spiegel_2015_ias15,
  author = {Rein, Hanno and Spiegel, David S.},
  title = {{IAS15}: A Fast, Adaptive, High-order Integrator for Gravitational Dynamics, Accurate to Machine Precision over a Billion Orbits},
  journal = {Monthly Notices of the Royal Astronomical Society},
  year = {2015},
  volume = {446},
  number = {2},
  pages = {1424--1437},
  doi = {10.1093/mnras/stu2164}
}

@article{galligan2005radiant,
  author = {Galligan, D. P. and Baggaley, W. J.},
  title = {The Radiant Distribution of {AMOR} Radar Meteors},
  journal = {Monthly Notices of the Royal Astronomical Society},
  year = {2005},
  volume = {359},
  number = {2},
  pages = {551--560},
  doi = {10.1111/j.1365-2966.2005.08918.x}
}

@article{hashimoto,
  author = {Hashimoto, Taishi and Saito, Akinori and Nishimura, Koji and Tsutsumi, Masaki and Sato, Kaoru and Sato, Toru},
  title = {First Incoherent Scatter Measurements and Adaptive Suppression of Field-Aligned Irregularities by the {PANSY} Radar at Syowa Station, Antarctic},
  journal = {Journal of Atmospheric and Oceanic Technology},
  year = {2019},
  volume = {36},
  number = {9},
  pages = {1881--1888},
  doi = {10.1175/JTECH-D-18-0175.1}
}

@article{janches2013initial,
  author = {Janches, Diego and Hormaechea, Jos{\'e} Luis and Brunini, C. and Hocking, W. and Fritts, D. C.},
  title = {An Initial Meteoroid Stream Survey in the Southern Hemisphere Using the Southern Argentina Agile Meteor Radar ({SAAMER})},
  journal = {Icarus},
  year = {2013},
  volume = {223},
  number = {2},
  pages = {677--683},
  doi = {10.1016/j.icarus.2012.12.018}
}

@article{janches2015southern,
  author = {Janches, D. and Close, S. and Hormaechea, J. L. and Swarnalingam, N. and Murphy, A. and O'Connor, D. and Vandepeer, B. and Fuller, B. and Fritts, D. C. and Brunini, C.},
  title = {The Southern Argentina Agile Meteor Radar Orbital System ({SAAMER-OS}): An Initial Sporadic Meteoroid Orbital Survey in the Southern Sky},
  journal = {The Astrophysical Journal},
  year = {2015},
  volume = {809},
  number = {1},
  eid = {36},
  doi = {10.1088/0004-637X/809/1/36}
}

@article{chau2007jicamarca_sources,
  author = {Chau, Jorge L. and Woodman, Ronald F. and Galindo, Freddy},
  title = {Sporadic Meteor Sources as Observed by the {Jicamarca} High-Power Large-Aperture {VHF} Radar},
  journal = {Icarus},
  year = {2007},
  volume = {188},
  number = {1},
  pages = {162--174},
  doi = {10.1016/j.icarus.2006.11.006},
  adsurl = {https://ui.adsabs.harvard.edu/abs/2007Icar..188..162C}
}

@article{jenniskens_vaubaillon_2010_capricornids,
  author = {Jenniskens, Peter and Vaubaillon, J{\'e}r{\'e}mie},
  title = {Minor Planet 2002 {EX}$_{12}$ (=169P/{NEAT}) and the Alpha Capricornid Shower},
  journal = {The Astronomical Journal},
  year = {2010},
  volume = {139},
  number = {5},
  pages = {1822--1830},
  doi = {10.1088/0004-6256/139/5/1822},
  adsurl = {https://ui.adsabs.harvard.edu/abs/2010AJ....139.1822J}
}

@article{brown_etal_2010_cmor_minor_showers,
  author = {Brown, P. and Wong, D. K. and Weryk, R. J. and Wiegert, P.},
  title = {A Meteoroid Stream Survey Using the Canadian Meteor Orbit Radar. II: Identification of Minor Showers Using a 3D Wavelet Transform},
  journal = {Icarus},
  year = {2010},
  volume = {207},
  pages = {66--81},
  doi = {10.1016/j.icarus.2009.11.015}
}

@book{jenniskens_2023_atlas,
  author = {Jenniskens, Peter},
  title = {Atlas of Earth's Meteor Showers},
  publisher = {Elsevier},
  year = {2023},
  edition = {1},
  isbn = {978-0-323-88447-1}
}

@article{brown_etal_2008_cmor_survey,
  author = {Brown, P. and Weryk, R. J. and Wong, D. K. and Jones, J.},
  title = {A Meteoroid Stream Survey Using the Canadian Meteor Orbit Radar},
  journal = {Icarus},
  year = {2008},
  volume = {195},
  pages = {317--339},
  doi = {10.1016/j.icarus.2007.12.002}
}

@article{campbell_brown_2008_sporadic_radiants,
  author = {Campbell-Brown, M. D.},
  title = {High Resolution Radiant Distribution and Orbits of Sporadic Radar Meteoroids},
  journal = {Icarus},
  year = {2008},
  volume = {196},
  number = {1},
  pages = {144--163},
  doi = {10.1016/j.icarus.2008.02.022}
}

@article{campbell_brown_wiegert_2009_toroidal,
  author = {Campbell-Brown, M. and Wiegert, P.},
  title = {Seasonal Variations in the North Toroidal Sporadic Meteor Source},
  journal = {Meteoritics \& Planetary Science},
  year = {2009},
  volume = {44},
  number = {12},
  pages = {1837--1848},
  doi = {10.1111/j.1945-5100.2009.tb01992.x}
}

@article{kastinenRadarAnalysisAlgorithm2022,
  title = {Radar Analysis Algorithm for Determining Meteor Head Echo Parameter Probability Distributions},
  author = {Kastinen, Daniel and Kero, Johan},
  year = 2022,
  month = sep,
  journal = {Monthly Notices of the Royal Astronomical Society},
  eprint = {https://academic.oup.com/mnras/advance-article-pdf/doi/10.1093/mnras/stac2727/46130771/stac2727.pdf},
  issn = {0035-8711},
  doi = {10.1093/mnras/stac2727},
}

@article{kastinen2020probabilistic,
  author = {Kastinen, Daniel and Kero, Johan},
  title = {Probabilistic Analysis of Ambiguities in Radar Echo Direction of Arrival from Meteors},
  journal = {Atmospheric Measurement Techniques},
  year = {2020},
  volume = {13},
  number = {12},
  pages = {6813--6835},
  doi = {10.5194/amt-13-6813-2020}
}

@inproceedings{kastinen_pansy,
  author = {Kastinen, D. and Kero, J. and Nishimura, K.},
  title = {Probabilistic Meteor Analysis and the First Head Echo Results from the {PANSY} Radar},
  booktitle = {2021 XXXIVth General Assembly and Scientific Symposium of the International Union of Radio Science ({URSI} {GASS})},
  year = {2021},
  pages = {1--4},
  publisher = {IEEE},
  doi = {10.23919/URSIGASS51995.2021.9560120}
}

@article{nishimura,
  author = {Nishimura, Koji and Kohma, Masashi and Sato, Kaoru and Sato, Toru},
  title = {Spectral Observation Theory and Beam Debroadening Algorithm for Atmospheric Radar},
  journal = {{IEEE} Transactions on Geoscience and Remote Sensing},
  year = {2020},
  volume = {58},
  number = {10},
  pages = {6767--6775},
  doi = {10.1109/TGRS.2020.2970200}
}

@article{picone2002nrlmsise,
  author = {Picone, J. M. and Hedin, A. E. and Drob, D. P. and Aikin, A. C.},
  title = {{NRLMSISE-00} Empirical Model of the Atmosphere: Statistical Comparisons and Scientific Issues},
  journal = {Journal of Geophysical Research: Space Physics},
  year = {2002},
  volume = {107},
  number = {A12},
  pages = {SIA 15-1--SIA 15-16},
  doi = {10.1029/2002JA009430}
}

@article{sato2014program,
  author = {Sato, Kaoru and Tsutsumi, Masaki and Sato, Toru and Nakamura, Takuji and Saito, Akinori and Tomikawa, Yoshihiro and Nishimura, Koji and Kohma, Masashi and Yamagishi, Hisao and Yamanouchi, Takashi},
  title = {Program of the Antarctic {Syowa MST/IS} Radar ({PANSY})},
  journal = {Journal of Atmospheric and Solar-Terrestrial Physics},
  year = {2014},
  volume = {118},
  pages = {2--15},
  doi = {10.1016/j.jastp.2013.08.022}
}

@article{vondrak2008chemical,
  author = {Vondrak, T. and Plane, J. M. C. and Broadley, S. and Janches, D.},
  title = {A Chemical Model of Meteoric Ablation},
  journal = {Atmospheric Chemistry and Physics},
  year = {2008},
  volume = {8},
  number = {23},
  pages = {7015--7031},
  doi = {10.5194/acp-8-7015-2008}
}

@article{janches2009differential,
  author = {Janches, D. and Dyrud, L. P. and Broadley, S. L. and Plane, J. M. C.},
  title = {First Observation of Micrometeoroid Differential Ablation in the Atmosphere},
  journal = {Geophysical Research Letters},
  year = {2009},
  volume = {36},
  pages = {L06101},
  doi = {10.1029/2009GL037389}
}

@article{bones2022organic,
  author = {Bones, D. L. and Carrillo-Sanchez, J. D. and Connell, S. D. A. and Kulak, A. N. and Mann, G. W. and Plane, J. M. C.},
  title = {Ablation Rates of Organic Compounds in Cosmic Dust and Resulting Changes in Mechanical Properties During Atmospheric Entry},
  journal = {Earth and Space Science},
  year = {2022},
  volume = {9},
  number = {4},
  pages = {e2021EA001884},
  doi = {10.1029/2021EA001884}
}

@article{huyghebaert2025altitude,
  author = {Huyghebaert, Devin and Vierinen, Juha and Kero, Johan and Mann, Ingrid and Latteck, Ralph and Kastinen, Daniel and Vaden, Sara and Chau, Jorge L.},
  title = {Examining the Altitude Dependence of Meteor Head Echo Plasma Distributions with {EISCAT} and {MAARSY}},
  journal = {Advances in Space Research},
  year = {2025},
  volume = {76},
  number = {4},
  pages = {2280--2294},
  doi = {10.1016/j.asr.2025.06.056}
}

@article{schult2013determination,
  author = {Schult, C. and Stober, G. and Chau, J. L. and Latteck, R.},
  title = {Determination of Meteor-head Echo Trajectories Using the Interferometric Capabilities of {MAARSY}},
  journal = {Annales Geophysicae},
  year = {2013},
  volume = {31},
  number = {10},
  pages = {1843--1851},
  doi = {10.5194/angeo-31-1843-2013}
}

@article{schult_2017_results,
  author = {Schult, Carsten and Stober, Gunter and Janches, Diego and Chau, Jorge L.},
  title = {Results of the First Continuous Meteor Head Echo Survey at Polar Latitudes},
  journal = {Icarus},
  volume = {297},
  pages = {1--13},
  year = {2017},
  doi = {10.1016/j.icarus.2017.06.019}
}

@article{schult_2018_meteoroid,
  author = {Schult, Carsten and Brown, Peter and Pokorn{\'y}, Petr and Stober, Gunter and Chau, Jorge L.},
  title = {A Meteoroid Stream Survey Using Meteor Head Echo Observations from the Middle Atmosphere {ALOMAR} Radar System ({MAARSY})},
  journal = {Icarus},
  volume = {309},
  pages = {177--186},
  year = {2018},
  doi = {10.1016/j.icarus.2018.02.032}
}

@article{kero_mu_catalogue,
  author = {Kero, J. and Szasz, C. and Nakamura, T. and Meisel, D. D. and Ueda, M. and Fujiwara, Y. and Terasawa, T. and Nishimura, K. and Watanabe, J.},
  title = {The 2009--2010 {MU} Radar Head Echo Observation Programme for Sporadic and Shower Meteors: Radiant Densities and Diurnal Rates},
  journal = {Monthly Notices of the Royal Astronomical Society},
  year = {2012},
  volume = {425},
  number = {1},
  pages = {135--146},
  doi = {10.1111/j.1365-2966.2012.21407.x},
  adsurl = {https://ui.adsabs.harvard.edu/abs/2012MNRAS.425..135K}
}

@article{kero_2012_meteor,
  author = {Kero, J. and Szasz, C. and Nakamura, T. and Terasawa, T. and Miyamoto, H. and Nishimura, K.},
  title = {A Meteor Head Echo Analysis Algorithm for the Lower {VHF} Band},
  journal = {Annales Geophysicae},
  volume = {30},
  number = {4},
  pages = {639--659},
  year = {2012},
  doi = {10.5194/angeo-30-639-2012},
  adsurl = {https://ui.adsabs.harvard.edu/abs/2012AnGeo..30..639K}
}

@article{nesvorny2010cometary,
  author = {Nesvorn{\'y}, David and Jenniskens, Peter and Levison, Harold F. and Bottke, William F. and Vokrouhlick{\'y}, David and Gounelle, Matthieu},
  title = {Cometary Origin of the Zodiacal Cloud and Carbonaceous Micrometeorites: Implications for Hot Debris Disks},
  journal = {The Astrophysical Journal},
  year = {2010},
  volume = {713},
  number = {2},
  pages = {816--836},
  doi = {10.1088/0004-637X/713/2/816}
}

@article{nesvorny2011dynamical,
  author = {Nesvorn{\'y}, David and Janches, Diego and Vokrouhlick{\'y}, David and Pokorn{\'y}, Petr and Bottke, William F. and Jenniskens, Peter},
  title = {Dynamical Model for the Zodiacal Cloud and Sporadic Meteors},
  journal = {The Astrophysical Journal},
  year = {2011},
  volume = {743},
  number = {2},
  pages = {129},
  doi = {10.1088/0004-637X/743/2/129}
}

@article{pokorny_2017_saamer_stream_survey,
  author = {Pokorn{\'y}, Petr and Janches, Diego and Brown, Peter G. and Hormaechea, Jose Luis and Weryk, Robert J.},
  title = {An Orbital Meteoroid Stream Survey Using the Southern Argentina Agile {ME}teor Radar ({SAAMER}) Based on a Wavelet Approach},
  journal = {Icarus},
  year = {2017},
  volume = {290},
  pages = {162--182},
  doi = {10.1016/j.icarus.2017.02.025}
}

@article{wiegert2009dynamical,
  author = {Wiegert, Paul and Vaubaillon, J{\'e}r{\'e}mie and Campbell-Brown, Margaret},
  title = {A Dynamical Model of the Sporadic Meteoroid Complex},
  journal = {Icarus},
  year = {2009},
  volume = {201},
  number = {1},
  pages = {295--310},
  doi = {10.1016/j.icarus.2008.12.030}
}

@article{jopek1993remarks,
  author = {Jopek, Tadeusz J.},
  title = {Remarks on the Meteor Orbital Similarity {D}-Criterion},
  journal = {Icarus},
  year = {1993},
  volume = {106},
  number = {2},
  pages = {603--607},
  doi = {10.1006/icar.1993.1195}
}

@article{sato2000mu,
  author = {Sato, Toru and Nakamura, Takuji and Nishimura, Koji},
  title = {Orbit Determination of Meteors Using the {MU} Radar},
  journal = {IEICE Transactions on Communications},
  year = {2000},
  volume = {E83-B},
  number = {9},
  pages = {1990--1995},
  month = sep,
  url = {https://globals.ieice.org/en_transactions/communications/10.1587/e83-b_9_1990/_p}
}

@article{soja2019imem2,
  author = {Soja, R. H. and Gr{\"u}n, E. and Strub, P. and Sommer, M. and Millinger, M. and Vaubaillon, J. and Alius, W. and Camodeca, G. and Hein, F. and Laskar, J. and Gastineau, M. and Fienga, A. and Schwarzkopf, G. J. and Herzog, J. and Gutsche, K. and Skuppin, N. and Srama, R.},
  title = {{IMEM2}: A Meteoroid Environment Model for the Inner Solar System},
  journal = {Astronomy \& Astrophysics},
  year = {2019},
  volume = {628},
  pages = {A109},
  doi = {10.1051/0004-6361/201834892}
}

@article{close_2007_meteor,
  author = {Close, S. and Brown, P. and Campbell-Brown, M. and Oppenheim, M. and Colestock, P.},
  title = {Meteor Head Echo Radar Data: Mass--Velocity Selection Effects},
  journal = {Icarus},
  year = {2007},
  volume = {186},
  number = {2},
  pages = {547--556},
  doi = {10.1016/j.icarus.2006.09.007}
}

@article{gorski2005healpix,
  author = {G{\'o}rski, Krzysztof M. and Hivon, Eric and Banday, Anthony J. and Wandelt, Benjamin D. and Hansen, Frode K. and Reinecke, Markus and Bartelmann, Matthias},
  title = {{HEALPix}: A Framework for High-Resolution Discretization and Fast Analysis of Data Distributed on the Sphere},
  journal = {The Astrophysical Journal},
  year = {2005},
  volume = {622},
  number = {2},
  pages = {759--771},
  doi = {10.1086/427976}
}

@article{renkwitz_2017_antenna,
  author = {Renkwitz, Toralf and Schult, Carsten and Latteck, Ralph},
  title = {{VHF} antenna pattern characterization by the observation of meteor head echoes},
  journal = {Atmospheric Measurement Techniques},
  volume = {10},
  number = {2},
  pages = {527--535},
  year = {2017},
  doi = {10.5194/amt-10-527-2017},
  url = {https://doi.org/10.5194/amt-10-527-2017},
}

@article{tarnecki_2021_meteoroid,
  author = {Tarnecki, L. K. and Marshall, R. A. and Stober, G. and Kero, J.},
  title = {Meteoroid Mass Estimation Based on Single-Frequency Radar Cross Section Measurements},
  journal = {Journal of Geophysical Research: Space Physics},
  volume = {126},
  number = {9},
  pages = {e2021JA029525},
  year = {2021},
  doi = {10.1029/2021JA029525},
  url = {https://doi.org/10.1029/2021JA029525},
}

@misc{juha_2026_21703650,
  author = {Vierinen, Juha},
  title = {The {PANSY} Meteor Head-echo Orbit Catalogue},
  month = jul,
  year = {2026},
  publisher = {Zenodo},
  doi = {10.5281/zenodo.21703650},
  url = {https://doi.org/10.5281/zenodo.21703650},
}

@Article{chau+woodman-2004,
AUTHOR = {Chau, J. L. and Woodman, R. F.},
TITLE = {Observations of meteor-head echoes using the Jicamarca 50MHz radar in interferometer mode},
JOURNAL = {Atmospheric Chemistry and Physics},
VOLUME = {4},
YEAR = {2004},
NUMBER = {2},
PAGES = {511--521},
URL = {https://acp.copernicus.org/articles/4/511/2004/},
DOI = {10.5194/acp-4-511-2004}
}

@article{chau+galindo-2008,
title = {First definitive observations of meteor shower particles using a high-power large-aperture radar},
journal = {Icarus},
volume = {194},
number = {1},
pages = {23-29},
year = {2008},
issn = {0019-1035},
doi = {https://doi.org/10.1016/j.icarus.2007.09.021},
url = {https://www.sciencedirect.com/science/article/pii/S001910350700471X},
author = {Jorge L. Chau and Freddy Galindo}
}

@article{chau_2019_empirical,
  author = {Chau, J. L. and Clahsen, M.},
  title = {Empirical Phase Calibration for Multistatic Specular Meteor Radars Using a Beamforming Approach},
  journal = {Radio Science},
  volume = {54},
  number = {1},
  pages = {60--71},
  year = {2019},
  doi = {10.1029/2018RS006741}
}

@article{janches2014saamer_head,
  author = {Janches, Diego and Hocking, Wayne and Pifko, Scott and Hormaechea, Jos{\'e} L. and Fritts, David C. and Brunini, Claudio and Michell, Robert and Samara, Marilia},
  title = {Interferometric Meteor Head Echo Observations Using the Southern Argentina Agile Meteor Radar},
  journal = {Journal of Geophysical Research: Space Physics},
  volume = {119},
  number = {3},
  pages = {2269--2287},
  year = {2014},
  doi = {10.1002/2013JA019241}
}

@article{michell2015saamer_head,
  author = {Michell, Robert G. and Janches, Diego and Samara, Marilia and Hormaechea, Jos{\'e} L. and Brunini, Claudio and Bibbo, Ignacio},
  title = {Simultaneous Optical and Radar Observations of Meteor Head-Echoes Utilizing {SAAMER}},
  journal = {Planetary and Space Science},
  volume = {118},
  pages = {95--101},
  year = {2015},
  doi = {10.1016/j.pss.2015.04.018}
}

@article{panka2021saamer_head,
  author = {Panka, Peter A. and Weryk, Robert J. and Bruzzone, Juan S. and Janches, Diego and Schult, Carsten and Stober, Gunter and Hormaechea, Jos{\'e} Luis},
  title = {An Improved Method to Measure Head Echoes Using a Meteor Radar},
  journal = {The Planetary Science Journal},
  volume = {2},
  number = {5},
  pages = {197},
  year = {2021},
  doi = {10.3847/PSJ/ac22b2}
}

@article{jenniskens2020workinglist,
  author = {Jenniskens, Peter and Jopek, Tadeusz J. and Janches, Diego and Hajdukov{\'a}, M{\'aria} and Kokhirova, Gulchehra I. and Rudawska, Regina},
  title = {On Removing Showers from the {IAU} Working List of Meteor Showers},
  journal = {Planetary and Space Science},
  volume = {182},
  pages = {104821},
  year = {2020},
  doi = {10.1016/j.pss.2019.104821}
}

@article{segon2024fornax,
  author = {{\v S}egon, Damir and Vida, Denis and Roggemans, Paul and Rollinson, David and Scott, James M.},
  title = {New Meteor Shower in Fornax},
  journal = {eMetN Meteor Journal},
  volume = {9},
  number = {5},
  pages = {279--285},
  year = {2024},
  url = {https://www.emeteornews.net/2024/07/28/new-meteor-shower-in-fornax/}
}

@article{roggemans2026m2024n1,
  author = {Roggemans, Paul and Vida, Denis and {\v S}egon, Damir and Scott, James M. and Wood, Jeff},
  title = {{M2024-N1} Activity Confirmed in 2025},
  journal = {eMetN Meteor Journal},
  volume = {11},
  number = {1},
  pages = {47--51},
  year = {2026},
  url = {https://www.emeteornews.net/2025/12/24/m2024-n1-activity-confirmed-in-2025/}
}

@article{greaves2026meteor_candidates,
  author = {Greaves, John},
  title = {Independent Confirmation of Meteor Shower Candidates from the Southern Argentine Agile Meteor Radar},
  journal = {eMetN Meteor Journal},
  volume = {11},
  number = {2},
  pages = {151--153},
  year = {2026},
  url = {https://www.emeteornews.net/wp-content/uploads/2026/02/eMetN2026_2.pdf}
}

@inproceedings{cook1973workinglist,
  author = {Cook, Allan F.},
  title = {A Working List of Meteor Streams},
  booktitle = {Evolutionary and Physical Properties of Meteoroids},
  editor = {Hemenway, Curtis L. and Millman, Peter M. and Cook, Allan F.},
  series = {NASA Special Publication},
  volume = {319},
  pages = {183--191},
  year = {1973},
  doi = {10.1017/S025292110004906X}
}

@misc{openai2025codex,
  author = {{OpenAI}},
  title = {Introducing Codex},
  year = {2025},
  month = may,
  url = {https://openai.com/index/introducing-codex/},
  note = {Accessed 2026 August 16}
}

@article{vishniac2023chatbots,
  author = {Vishniac, Ethan T.},
  title = {Editorial: On the Use of Chatbots in Writing Scientific Manuscripts},
  journal = {Bulletin of the AAS},
  year = {2023},
  volume = {55},
  number = {1},
  doi = {10.3847/25c2cfeb.c3619710}
}

\end{document}